\documentclass[a4paper,oneside,final,notitlepage,onecolumn,12pt]{article}
\usepackage{amsfonts}
\usepackage{epsf}
\usepackage{graphicx}
\usepackage{amssymb,eso-pic}
\usepackage{latexsym}
\usepackage{tabularx}
\usepackage{amsxtra}
\usepackage{hyperref}
\usepackage{t1enc}
\usepackage{amsmath}
\usepackage[T1]{fontenc}
\usepackage{amsmath,accents}
\usepackage{authblk}

\usepackage{framed}
\usepackage{enumerate}
\usepackage{bm}
\usepackage{mathrsfs}
\usepackage{amsthm}
\usepackage{latexsym}
\usepackage{tikz}
\usepackage{mathtools}
\DeclareGraphicsExtensions{.jpg,.jpeg,.pdf,.png,.eps}
\usepackage{cite}
\usepackage{caption}
\usepackage{subcaption}
\usepackage{tabularx}
\usepackage[abs]{overpic}
\usepackage{multirow}

\usepackage[utf8]{inputenc}
\usepackage{ifpdf,epsfig,array,amsmath,amssymb}

\usepackage{psfrag}
\psfrag{na}{\footnotesize $n^a$}
\psfrag{ta}{\footnotesize $\tau^a$}
\psfrag{vna}{\footnotesize $\check{n}^a$}
\psfrag{vNa}{\footnotesize $\check{N}^a$}
\psfrag{scrW}{\footnotesize $\mathscr{W}_{\rho_0}$}
\psfrag{t=t1}{\footnotesize $\Sigma_{\tau_1}$}
\psfrag{t=t2}{\footnotesize $\Sigma_{\tau_2}$}
\psfrag{r=all}{\footnotesize $\rho=const$}
\psfrag{hna}{\footnotesize $\hat{n}^a$}
\psfrag{tna}{\footnotesize $\tilde{n}^a$}
\psfrag{hab,Kab}{\footnotesize $h_{ab}, K_{ab}$}
\psfrag{thab,tKab}{\footnotesize $\tilde h_{ab}, \tilde K_{ab}$}
\psfrag{hhab,hKab}{\footnotesize $\widehat h_{ab}, \widehat K_{ab}$}
\psfrag{chab,cKab}{\footnotesize $\check{h}_{ab}, \check{K}_{ab}$}

\usepackage[normalem]{ulem}
\usepackage{color}
\definecolor{blue}{rgb}{0,0,1}
\definecolor{red}{rgb}{1,0,0}

\DeclareFontFamily{OT1}{rsfs}{} \DeclareFontShape{OT1}{rsfs}{m}{n}{
<-7> rsfs5 <7-10> rsfs7 <10-> rsfs10}{}
\DeclareMathAlphabet{\mathscr}{OT1}{rsfs}{m}{n}

\def\scri{{\mathscr I}}
\def\scrip{\scri^{+}}%

\def\sc{{\hskip 3.5pt {{}^{{}^{{}_{{}_{\bowtie}}}}} \kern -8.pt{}}}
\def\SC{{\hskip 3.5pt {{}^{{}^{{}^{{}_{{}_{\bowtie}}}}}} \kern -10.5pt{}}}

\newcommand{\dd}{\mathrm{d}}
\newcommand{\diff}{\partial}

\definecolor{green1}{HTML}{008000}
\definecolor{green2}{HTML}{00ee00}
\definecolor{green3}{HTML}{00cd00}
\definecolor{olive-green}{HTML}{556B2F}
\definecolor{chartreuse}{HTML}{66cd00}
\definecolor{DGREEN}{rgb}{0,0.65,0.65}

\newcommand{\diffr}[1]{\textcolor{orange}{\textbf{#1}}}
\newcommand{\diffg}[1]{\textcolor{DGREEN}{\textbf{#1}}}

\usepackage[normalem]{ulem}
\usepackage{color}

\newcounter{mnotecount}

\newcommand{\mnotex}[1]
{\protect{\stepcounter{mnotecount}}$^{\mbox{\footnotesize $\bullet$\themnotecount}}$
	\marginpar{\color{red}
		\raggedright\tiny\em
		$\!\!\!\!\!\!\,\bullet$\themnotecount: #1} }
\newcommand{\Wj}[6]{\left(\begin{array}{ccc}#1&#2&#3\\#4&#5&#6\end{array}\right)}
\begin{document}

\title{\textbf{Numerical investigation of the dynamics of linear spin $s$ fields on a Kerr background:
	 \\ Late time tails of spin $s=\pm 1,\pm 2$ fields
	}}

\author[,1]{K\'aroly Csuk\'as \footnote{E-mail address:{\tt csukas.karoly@wigner.mta.hu}}}

\author[,1,2]{Istv\'an R\'acz \footnote{E-mail address:{\tt racz.istvan@wigner.mta.hu}}}

\author[,1]{G\'abor Zsolt T\'oth \footnote{E-mail address:{\tt toth.gabor.zsolt@wigner.mta.hu}}}

\affil[1]{Wigner RCP, H-1121 Budapest, Konkoly Thege Mikl\'{o}s \'{u}t  29-33, Hungary}

\affil[2]{Faculty of Physics, University of Warsaw, Ludwika Pasteura 5, 02-093 Warsaw, Poland}

\maketitle

\begin{abstract}
The time evolution of linear fields of spin $s=\pm 1$ and $s=\pm 2$ on Kerr black hole spacetimes are investigated by solving the homogeneous Teukolsky equation numerically. The applied numerical setup is based on a combination of conformal compactification and the hyperbolic initial value problem. The evolved basic variables are expanded in terms of spin-weighted spherical harmonics, which allows us to evaluate all angular derivatives analytically, whereas the evolution of the expansion coefficients in the time-radial section is determined by applying the method of lines implemented in a fourth order accurate finite differencing stencil. Concerning the initialization, in all of our investigations, single mode excitations---either static or purely dynamical-type initial data---are applied. Within this setup the late-time tail behavior is investigated. Because of the applied conformal compactification, the asymptotic decay rates are determined at three characteristic locations---in the  domain of outer communication, at the event horizon, and at future null infinity---simultaneously.
A recently introduced new type of ``energy'' and ``angular momentum'' balance relations are also applied in order to demonstrate the feasibility and robustness of the developed numerical schema and also to verify the proper implementation of the underlying mathematical model.
\end{abstract}



\section{Introduction}\label{introduction}
\setcounter{equation}{0}

The study of the long-time evolution of various linear fields on given black hole backgrounds has served for decades as a preparation for the more involved study of linear and possibly nonlinear stability of the background black hole solutions themselves. In the last couple of years considerable progress has been made concerning the linear stability of the Kerr solution (see, e.g.~\cite{dafermos-et-all:2017,lars-et-all:2019} and references therein). Notably, even these analytic results rest on various technical assumptions which apparently all boil down to the long-time behavior of linear spin $s$ fields satisfying the Teukolsky master equation \cite{dafermos-et-all:2017,lars-et-all:2019}. This provides immediate motivation for a thorough investigation of the asymptotic in time behavior of solutions to the Teukolsky equation on a Kerr black hole background.

\medskip

This paper reports about our findings concerning numerical investigations of the dynamics of electromagnetic and gravitational perturbations on a Kerr background. Our aim was to carry out comprehensive investigations of the time evolution of linear spin $s$ fields, in particular, to study their tail behavior. This is done by solving the homogeneous Teukolsky master equation for generic linear spin $s=\pm 1$ and $s=\pm 2$ fields numerically.
Since previous numerical studies focused mostly on axisymmetric configurations,
we aimed to provide a more detailed study of nonaxisymmetric configurations.
In doing so, the analytic framework was chosen such that it incorporates both techniques of conformal compactification and the hyperboloidal initial value problem. The time slices in the latter were chosen to be horizon penetrating, which allows us to determine  the decay rates at the three characteristic locations simultaneously: at the black hole event horizon, in the domain of outer communication, and at future null infinity. In addition, the applied mathematical setup also makes it possible to use a spherical spectral representation of all the basic variables. This is so as they are expanded in terms of spin-weighted spherical harmonics based on the foliation of the Kerr background by topological two-spheres, which in practice are the Boyer-Lindquist $t=const$ and $r=const$ ellipsoids. As all angular derivatives can be given then either in terms of the ``eth'' and ``ethbar'', $\eth$ and $\overline{\eth}$, operators or by Lie derivatives with respect to the axial symmetry of the Kerr background, all angular derivatives are evaluated analytically. This, in turn, guarantees that the pertinent multipole expansion coefficients---as elements of a large set of coupled scalar fields---get to be subject of an involved but otherwise only  $(1+1)$-dimensional time-evolution problem.
Notably, even the case of the most generic linear spin $s$ fields can be tested within the very same mathematical and numerical setup which, in turn, allows us to treat all the nonaxisymmetric configurations as well. Though in practice, our attention was restricted to the range $m=0,\pm 1,\pm 2$ of the azimuthal parameter, in principle, one could investigate the full spectra of nonaxisymmetric configurations within the chosen framework.

\medskip

In checking and interpreting our numerical findings it was very informative and supportive to compare them to both analytic and numerical results prior to our investigations. On a Kerr background, the first systematic analytic estimates on the decay exponents were derived by  Barack and Ori \cite{Barack:1999ma,Barack:1999st}. They developed a method capable of studying the evolution of linear spin $s=0$, $s=\pm1$, and $s=\pm2$ in the time domain. By applying this they discovered an interesting phenomenon. Namely, for axially symmetric configurations, the decay rates at the event horizon are larger by $1$ for spin $s>0$ with respect to those relevant for $s<0$ \cite{Barack:1999ya}.
By applying the same method, they could also study the behavior of linear perturbations while approaching the Cauchy or inner horizon of the Kerr background \cite{Ori:2001pc,Ori:1999nc}. The alternative analytic investigations by Hod were carried out in the frequency domain \cite{Hod:1999ci,Hod:1999rx,Hod:2000fh}. In \cite{Casals:2015nja,Casals:2016soq} Casals et al.\ developed a technique to further improve on the results of Hod. They presented a method to include higher order terms in the low-frequency expansion, which is necessary for computing self-force.
Some complementary analytic studies can also be found in the Appendix of \cite{Harms:2013ib}.
Note that in Ref. \cite{Harms:2013ib}, Harms et al. also report impressive and comprehensive numerical investigations concerning the long-time evolution of axially symmetric linear spin $s$ fields on a Kerr background.
It is important to emphasize that the initial data we apply are always of pure mode excitation, which allows us to study the long-time behavior of various higher mode excitations. It is of crucial importance then that the analytic and numerical investigations carried out in \cite{Hod:2000fh,Harms:2013ib} are also based on the use of single-mode-excitation-type initial data. In this respect, from among all the aforementioned excellent investigations, the ones reported in \cite{Hod:2000fh,Harms:2013ib} will be at the center of our interest.
Most of the predictions made in these studies are confirmed by our investigations. Nevertheless, as there were some slight disagreements even between the predictions of these two sets of investigations, in our studies particular attention was given to clear up of the corresponding cases. Notably, our numerical results led us to the conclusion that, though only in some very special subcases, neither of the former predictions was satisfactory: i.e.,~they did not give the correct value for the decay exponents.

\medskip

In coming up with a firm statement of the above type,
it is extremely important to guarantee the self-consistency of the underlying numerical results. It is indeed critical to verify that our findings are not simply numerical artifacts of the applied method, but they are rooted in the true nature of the investigated fields. For this reason, we implemented---in addition to the conventional convergence rate checks---the recently introduced \cite{Toth:2018ybm} new type of ``energy'' and ``angular momentum'' balance relations to verify both the proper implementation of the underlying mathematical model and the feasibility and robustness of the developed numerical schema.

\bigskip

Before proceeding in presenting our main results, it appears to be rewarding to recall that linear spin $s=\pm 1$ or $s=\pm 2$ fields satisfying the homogeneous Teukolsky master equations arise in a straightforward way in studying evolution of source-free electromagnetic fields in Maxwell theory or in that of linear metric perturbations on a Kerr background \cite{Teukolsky:1972my,Teukolsky:1973ha}. To see this, e.g.~in the electromagnetic case, recall first that in applying the Newman-Penrose formalism \cite{NP:1962np}, one starts by fixing a complex null tetrad $\{l^a, n^a, m^a, \overline{m}{}^a\}$
comprising at each point two real $l^a$ and $n^a$ and two complex $m^a$ and $\overline{m}{}^a$ null vectors such that their only nonzero inner products are  $l^an_a=-m^a\overline{m}_a=1$. We fix the remaining guage freedom by using the Kinnersley tetrad. The algebraically independent components of the Faraday tensor $F_{ab}$ can always be represented by the three complex Maxwell scalar fields
\begin{align}\label{eq: Maxwell_phi}
	\phi_0=&\ F_{ab}l^am^b\,,\\
	\phi_1=&\ \frac{1}{2}\Big(F_{ab}l^an^b+F_{ab}\overline{m}^am^b\Big)\,,\\
	\phi_2=&\ F_{ab}\overline{m}^an^b\,.
\end{align}
Two of these, $\phi_0$ and $\phi_2$, represent the outgoing and ingoing radiations, respectively, whereas $\phi_1$ stores the Coulombic part of the electromagnetic field.

If $F_{ab}$ satisfies the source-free Maxwell equations on a Kerr black hole background,
then some of the one-order-higher wave equations---which can be deduced for the individual Maxwell scalar components $\phi_0$ and $\phi_2$, respectively---decouple. In particular, the fields $\psi^{(+1)}$ and $\psi^{(-1)}$ defined as
\begin{align}
\psi^{(+1)}=&\ \phi_0\,,\\
\psi^{(-1)}=&\ (\Psi_2)^{-2/3}\cdot\phi_2\,
\end{align}
satisfy the homogeneous Teukolsky master equation \eqref{eq:TMEBL} with spin $s=+1$ and $s=-1$, respectively \cite{Teukolsky:1972my}. Here, $\Psi_2$ is the only nonvanishing and gauge invariant Weyl-scalar component on a Kerr background which has the following form in Boyer-Lindquist coordinates using the Kinnersley tetrad:
\begin{equation}
	\Psi_2=-\frac{M}{(r-\mathrm{i}\, a \cos\vartheta)^3}.
\end{equation}

\medskip

By a completely analogous argument, it can also be verified \cite{Teukolsky:1972my,Teukolsky:1973ha} that in the case of linear metric perturbations, the specific contractions
\begin{equation}
\Psi_0=-C_{abcd}\,l^am^bl^cm^d \qquad {\textrm{and}} \qquad \Psi_4=-C_{abcd}\,n^a\overline{m}{}^{\,b}n^c\overline{m}{}^{\,d}
\end{equation}
of the Weyl tensor $C_{abcd}$---which is also equal to the curvature tensor in the case of vacuum spacetimes---and elements of the  complex null tetrad $\{l^a, n^a, m^a, \overline{m}{}^a\}$ are distinguished like $\phi_0$ and $\phi_2$ were in the electromagnetic case. Indeed, the linear wave equations---these can be deduced from the Bianchi identity, and they are relevant for the linearly perturbed configurations---really do decouple, and the fields
\begin{align}
\psi^{(+2)}=&\ \Psi_0\,,\\
\psi^{(-2)}=&\ (\Psi_2)^{-4/3}\cdot\Psi_4
\end{align}
satisfy the homogeneous Teukolsky master equation \eqref{eq:TMEBL} relevant for these spin $s=+2$ and $s=-2$ fields, respectively \cite{Teukolsky:1972my}.

\bigskip

This paper is organized as follows. In Sec. \ref{sec: lin-fields} first the analytic framework is introduced. This part (see Sec. \ref{subsec: regularization}) is to set up the form of the wave equation to be solved once the conformal compactification and suitable regularizations of the basic variables have been carried out. Some of the details of the new type of energy and angular momentum balance relations are also discussed in Sec. \ref{subsec: conserved-currents}. More details on the analytic and numerical setup are given in Sec. \ref{sec: more-analytic} containing the discussion of multipole expansions, the choice made for the initialization of time evolution, the determination of the decay rates and a summary of the results prior to ours.
Sec. \ref{sec: numerical-results} presents all of our numerical findings. We start by discussing axially symmetric configurations in Sec. \ref{subsec: axially-symm-conf} which is followed by a detailed investigation of nonaxisymmetric configurations in Sec. \ref{subsec: non-axially-symm-conf}. A thorough discussion of the use of the energy and angular momentum balance relations, as well as their use in verifying the convergence properties of the applied numerical implementation is given in Sec. \ref{subsec: energy-balance-relations}.
The discussions are completed by our final remarks in Sec. \ref{sec: final-results}. The paper is closed by several appendixes providing useful details on the applied analytic and numerical settings.

\section{Linear fields of spin $s$ on a Kerr background}\label{sec: lin-fields}

This section provides all the details related to the applied analytic setup.
In addition to the various forms of the Teukolsky master equation,
the energy and angular momentum balance relations
will also be discussed briefly.
This section also lays down the mathematical groundwork that will be applied in later papers of this series.

\subsection{Teukolsky equation}

The metric of the Kerr background in Boyer-Lindquist coordinates $(t,r,\vartheta,\phi)$ can be given by the line element
\begin{multline}
 (\dd s)^2=\left(1-\frac{2Mr}{\Sigma}\right)(\dd t)^2+\frac{4arM}{\Sigma}\sin^2\vartheta\,\dd t\dd\phi\\-\frac{\Sigma}{\Delta}\,(\dd r)^2-\Sigma\,(\dd\vartheta)^2-\frac{(r^2+a^2)^2-a^2\Delta\sin^2\vartheta}{\Sigma}\sin^2\vartheta\,(\dd\phi)^2,
\end{multline}
where $\Sigma=r^2+a^2\cos^2\vartheta$ and $\Delta=r^2-2Mr+a^2$, whereas $M$ and $a$ are the mass and the angular momentum per unit mass parameters of the Kerr black hole.

A linear sourceless field of spin $s$---where $s$ is integer or half-integer---on a Kerr background using the Kinnersley tetrad
is known to be subject to the homogeneous Teukolsky master equation \cite{Teukolsky:1972my}
\begin{multline}
	\label{eq:TMEBL}
	\left[\frac{(r^2+a^2)^2}{\Delta}-a^2\sin^2\vartheta\right]\frac{\diff^2\psi^{(s)}}{\diff t^2}+\frac{4Mar}{\Delta}\frac{\diff^2\psi^{(s)}}{\diff t\diff\phi}+\left[\frac{a^2}{\Delta}-\frac{1}{\sin^2\vartheta}\right]\frac{\diff^2\psi^{(s)}}{\diff\phi^2}-\\
	-\Delta^{-s}\frac{\diff}{\diff r}\left(\Delta^{s+1}\frac{\diff\psi^{(s)}}{\diff r}\right)-\frac{1}{\sin\vartheta}\frac{\diff}{\diff\vartheta}\left(\sin\vartheta\frac{\diff\psi^{(s)}}{\diff\vartheta}\right)-2s\left[\frac{a(r-M)}{\Delta}+\frac{\mathrm{i}\cos\vartheta}{\sin^2\vartheta}\right]\frac{\diff\psi^{(s)}}{\diff\phi}-\\-2s\left[\frac{M(r^2-a^2)}{\Delta}-r-\mathrm{i} a\cos\vartheta\right]\frac{\diff\psi^{(s)}}{\diff t}+(s^2\cot^2\vartheta-s)\psi^{(s)}=0.
\end{multline}

A very simple covariant form of the Teukolsky master equation was introduced by Bini et al.\
in \cite{Bini:2002jx} which allows us to write \eqref{eq:TMEBL} in the compact form
\begin{equation}
\label{eq:TMEBini}
\left[\,\left(\nabla^a+s\,\Gamma^a\right)\,\left(\nabla_a+s\,\Gamma_a\right)-4\,s^2\,\Psi_2\,\right]\,\psi^{(s)}=0\,,
\end{equation}
where the components of the ``connection vector'' $\Gamma^a$ are \cite{Bini:2002jx}
\begin{align}
\Gamma^t&=-\frac{1}{\Sigma}\left[\frac{M(r^2-a^2)}{\Delta}-(r+\mathrm{i}\, a \cos\vartheta)\right],\\
\Gamma^r&=-\frac{1}{\Sigma}\,\big(r-M\big),\\
\Gamma^\vartheta&=0,\\
\Gamma^\phi&=-\frac{1}{\Sigma}\left[\frac{a(r-M)}{\Delta}+\mathrm{i}\,\frac{\cos\vartheta}{\sin^2\vartheta}\right]\,.
\end{align}

\subsection{Regularization of the basic variables}\label{subsec: regularization}

The solutions to the homogeneous Teukolsky master equation are known to get singular either in the $\Delta\rightarrow 0$ or $r \rightarrow \infty$ limit.
In order to regularize by suitable rescaling of the field variables
and also to get the desired conformal compactification of the Kerr background, new coordinates are introduced.
This is done by following the proposal in \cite{Racz:2011qu} in two succeeding steps.
First, the Boyer-Lindquist coordinates $(t,r,\vartheta,\phi)$ are replaced by the ingoing Kerr coordinates $(\tau,r,\vartheta,\varphi)$, where the new time and azimuthal coordinates $\tau$ and $\varphi$ are
defined via the relations
\begin{align}
\label{eq:tr:ki}
\tau&=t-r+\int\dd r\frac{r^2+a^2}{\Delta},\\
\varphi&=\phi+\int\dd r\frac{a}{\Delta}\,.
\end{align}
Note that the $\tau=const$ hypersurfaces, as indicated in Fig.1.~in \cite{Racz:2011qu},
are horizon penetrating, whereas they all tend to spacelike infinity in the $r \rightarrow \infty$ limit.

In the second step, new coordinates $(T,R)$ replacing $(\tau,r)$ are introduced by the implicit relations
\begin{align}
\label{eq:tr:rt}
\tau&=T+\frac{1+R^2}{1-R^2}-4M\log(|1-R^2|)\,,\\
r&=\frac{2R}{1-R^2}\,.
\end{align}
The most important advantage of the application of these new coordinates is
that they allow the use of a conformal compactification of the Kerr spacetime such that
future null infinity ($\mathscr{I}^+$) gets to be represented by the $R=1$ hypersurface
and also that all the $T=const$ hypersurfaces are such that
they are both horizon penetrating and intersecting $\mathscr{I}^+$ in regular spherical cuts at $R=1$.

In order to get the desired suitably regularized basic field variables on the Kerr background
$\psi^{(s)}$ is replaced by $\Phi^{(s)}$ defined as
\begin{equation}
\Phi^{(s)}(T,R,\vartheta,\varphi)=\big[\,r(R)\cdot\Delta^s(R)\,\big]\cdot\psi^{(s)}(T,R,\vartheta,\varphi)\,.
\end{equation}

Once all the foregoing steps have been performed, the homogeneous Teukolsky master equation \eqref{eq:TMEBini} can be seen to take the form
\begin{align}
\label{eq:teurt}
&\diff_{TT}\Phi^{(s)}=\frac{1}{\mathscr{A}+\mathscr{B}\cdot Y_2^0}\,\Big[c_{RR}\cdot\diff_{RR}\Phi^{(s)}+c_{TR}\cdot\diff_{TR}\Phi^{(s)}+c_{T\varphi}\cdot\diff_{T\varphi}\Phi^{(s)}+c_{R\varphi}\cdot\diff_{R\varphi}\Phi^{(s)} \nonumber\\
&+c_{\vartheta\vartheta}\cdot\overline{\eth}\eth\,\Phi^{(s)}+c_T\cdot\diff_T\Phi^{(s)}+\mathrm{i}\,c_{Ty} \,Y_1^0\cdot\diff_T\Phi^{(s)}+c_R\cdot\diff_R\Phi^{(s)}+c_\varphi\cdot\diff_\varphi\Phi^{(s)}+c_0\cdot\Phi^{(s)}\Big]\,,
\end{align}
where $Y_1^0$ and $Y_2^0$ stand for the zero spin-weight spherical harmonics with $\ell=1,2$ and $m=0$,
whereas the explicit form of the involved $R$-dependent coefficients is given in Appendix \ref{app:TME_coeff}.
Note also that by making use of the $\eth$ and $\overline{\eth}$ operators acting on a spin $s$ field $f$ as
\begin{align}
\eth f&=-\sin^s\vartheta\left(\partial_\vartheta+\frac{\mathrm{i}}{\sin\vartheta}\partial_\varphi\right)(\sin^{-s}\!\vartheta \cdot f),\\
\overline{\eth} f&=-\sin^{-s}\vartheta\left(\partial_\vartheta-\frac{\mathrm{i}}{\sin\vartheta}\partial_\varphi\right)(\sin^{s}\!\vartheta \cdot f)\,,
\end{align}
all the $\vartheta$ derivatives present in the Laplace-Beltrami operator can be incorporated into a single operator $\overline{\eth}\eth$ via the relation
\begin{equation}
\overline{\eth}\eth f=\diff_{\vartheta\vartheta} f+\cot\vartheta\,\diff_\vartheta f+\frac{1}{\sin^2\vartheta}\,\diff_{\varphi\varphi} f+2\,\mathrm{i}\,s\,\frac{\cot\vartheta}{\sin\vartheta}\,\partial_\varphi f+s\,(1-s\cot^2\vartheta)f\,.
\end{equation}
This provides us considerable simplification and enhances the accuracy of our numerical integrator significantly,
as our approach is based on the use of multipole expansions of the basic variables
in terms of spin-weighted spherical harmonics, and all the angular derivatives in \eqref{eq:teurt} can be evaluated analytically.
(More details on the use of spin-weighted spherical harmonics and their relations to the $\eth$ and $\overline{\eth}$ operators
can be found in Appendix \ref{app: eth-bareth}.)

\subsection{Conserved currents}\label{subsec: conserved-currents}

It is known that there is no way to construct a Lagrangian
out of a single spin $s$ variable $\psi^{(s)}$ and its first order derivatives
such that one could get \eqref{eq:TMEBini} as the corresponding Euler-Lagrange equation for $\psi^{(s)}$.
Nevertheless, as pointed out in \cite{Toth:2018ybm} recently,
it is possible to associate a meaningful Lagrangian to a pair of spin $s$ and $-s$ fields via the relation
\begin{equation}\label{eq: Lagrangian}
\mathcal{L}=-(\nabla^a-s\Gamma^a)\psi^{(-s)}(\nabla_a+s\Gamma_a)\psi^{(s)}-4\,s^2\Psi_2\,\psi^{(s)}\psi^{(-s)}\,.
\end{equation}
It was also shown in \cite{Toth:2018ybm} that by making use of some suitable infinitesimal transformations
of the form $\psi^{(\pm s)}\to \psi^{(\pm s)}- \varsigma h^a\diff_a\psi^{(\pm s)}$ of the Lagrangian in \eqref{eq: Lagrangian},
canonical conserved Noether currents can also be associated with a pair of spin $s$ and $-s$ fields.
Note that the involved fields may be completely independent.
The only requirement for the conservation of the currents is that they both satisfy their respective Teukolsky master equations.
In particular, as the Lagrangian in \eqref{eq: Lagrangian} is invariant with respect to
the one-parameter group of diffeomorphisms induced by the Killing vector fields
$h^a=T^a=(\diff_T)^a$ and $h^a=\varphi^a=(\diff_\varphi)^a$ (which are also coordinate basis fields),
the corresponding infinitesimal transformations
endow us with the canonical energy- and angular-momentum-type currents defined as
\begin{align}
\label{eq:curr1}
E^a&=(\nabla^a-s\Gamma^a)\psi^{(-s)}T^b\diff_b\psi^{(s)}+(\nabla^a+s\Gamma^a)\psi^{(s)}T^b\diff_b\psi^{(-s)}+ T^a\mathcal{L}\,,\\
\label{eq:curr2}
J^a&=(\nabla^a-s\Gamma^a)\psi^{(-s)}\varphi^b\diff_b\psi^{(s)}+(\nabla^a+s\Gamma^a)\psi^{(s)}\varphi^b\diff_b\psi^{(-s)}+\varphi^a\mathcal{L}\,.
\end{align}
Notice that for $s=0$, (\ref{eq:curr1}) and (\ref{eq:curr2}) produce the well-known energy and angular momentum of the massless complex scalar field if $\psi^{(-s)}\big|_{s=0}=\overline{\psi^{(s)}}\big|_{s=0}$.

As the covariant divergence of these currents vanish, the balance relations
\begin{align}
\int_\Omega\nabla_aE^a=\int_{\diff\Omega}n_aE^a&=0\,,\label{eq: balance-E}\\
\int_\Omega\nabla_aJ^a=\int_{\diff\Omega}n_aJ^a&=0\label{eq: balance-J}
\end{align}
hold. The spacetime domain of integration $\Omega$ in all of our applications
is chosen to be the rectangular coordinate domain in $(T,R)$ such that
it is bounded by some initial and final time slices $T=T_i$ and $T=T_f$
and by some inner and outer timelike or null cylinders given by the $R=R_{in}$ and $R=R_{out}$ hypersurfaces, respectively.
In particular, denoting by $n_a^{(T)}=(\dd T)_a/\sqrt{g^{TT}}$ and $n_a^{(R)}=(\dd R)_a/\sqrt{-g^{RR}}$
the respective normals of the $T=const$ and $R=const$ hypersurfaces
and by $h_T$ and $h_R$ the determinant of the restriction of the metric to these hypersurfaces,
the energy balance relation can be given as
\begin{multline}
0=\int_{T=T_i}n_a^{(T)}E^a\sqrt{|h_T|}\,\,\dd R\,\dd\vartheta\,\dd\varphi+\int_{R=R_{in}}n_a^{(R)}E^a\sqrt{|h_R|}\,\,\dd T\,\dd\vartheta\,\dd\varphi\\
-\int_{T=T_f}n_a^{(T)}E^a\sqrt{|h_T|}\,\,\dd R\,\dd\vartheta\,\dd\varphi-\int_{R=R_{out}}n_a^{(R)}E^a\sqrt{|h_R|}\,\dd T\,\dd\vartheta\,\dd\varphi\,. \label{eq: balance}
\end{multline}
A completely analogous balance relation can be derived for the angular momentum simply by replacing $E^a$ in \eqref{eq: balance} with $J^a$.
The explicit form of the integrands $n_a^{(T)}E^a\sqrt{|h_T|}$, $n_a^{(R)}E^a\sqrt{|h_R|}$,
$n_a^{(T)}J^a\sqrt{|h_T|}$, and $n_a^{(R)}J^a\sqrt{|h_R|}$
involved in the energy and angular momentum balance relations can be found in Appendix \ref{app:curr}.

\section{More on the analytic and numerical setup}\label{sec: more-analytic}

The discussions in this section will remain valid for any linear field
of spin $s$ satisfying the homogeneous Teukolsky master equation on a Kerr black hole background.

\subsection{Multipole expansions}

In solving \eqref{eq:teurt}, our basic variables $\Phi^{(s)}$ are expanded in terms of spin-weight $s$ spherical harmonics ${}_s{Y_\ell}{}^m$ as
\begin{equation}\label{eq: mult-exp}
\Phi^{(s)}(T,R,\vartheta,\varphi) = \sum_{\ell=|s|}^{\ell_{max}}\sum_{m=-\ell}^{\ell}\phi_\ell{}^{m}(T,R)\cdot {}_s{Y_\ell}{}^m(\vartheta,\varphi)\,.
\end{equation}
In this way, \eqref{eq:teurt} becomes a set of coupled $(1+1)$-dimensional linear wave equations for the expansion coefficients $\phi_\ell{}^{m}(T,R)$, whereas all the angular derivatives are evaluated analytically by making use of the $\eth$ and $\overline{\eth}$ operators (more details on spin-weighted spherical harmonics and the $\eth$, $\overline{\eth}$ operators can be found in Appendix \ref{app: eth-bareth}). The summation goes from $\ell=|s|$ to some $\ell=\ell_{max}$ value, which is chosen to be suitably large in order to keep the truncation error tolerably small (which, in practice, corresponds to numerical precision).

Note that in the frequency domain analysis, the eigenfunctions of the angular part of Teukolsky master equation (TME) are spin-weighted spheroidal harmonics instead of spin-weighted spherical harmonics. In the work of Casals et al. \cite{Casals:2016soq}, the authors demonstrated that expanding in terms of spherical harmonics instead of spheroidal ones results in branch cuts in the complex Green's function. However, in the integral, the contributions of these extra cuts cancel out, so in the end the Green's function will be the same, and it is safe to use spin-weighted spherical harmonics as a basis. Since the spheroidal harmonics themselves can be expanded in terms of spherical harmonics, at least the decay rate of the slowest decaying modes is the same regardless of the basis of expansion.

By applying standard order reduction techniques---by introducing
$(\phi_{T}){}_\ell{}^m=\partial_T\phi_\ell{}^m$ and $(\phi_{R}){}_\ell{}^m=\partial_R\phi_\ell{}^m$
as additional dependent variables---a first order strongly hyperbolic system
is introduced for a vector variable that is composed of
the multipole expansion coefficients.
These equations were evolved in our numerical code by applying the method of lines
in a fourth order Runge-Kutta integrator and also using a sixth order dissipation term
for suppressing high-frequency spurious solutions \cite{1995tpdm}.

\subsection{The applied initial data}\label{subsec: initial-data}

In solving  \eqref{eq:teurt} in any time-evolution scheme it is necessary to initialize $\Phi^{(s)}$ by specifying, on some $T=T_0\,(\in\mathbb{R})$ initial data surface, a pair of functions $(\phi^{(s)},\phi_T^{(s)})$ such that  $\phi^{(s)}=\Phi^{(s)}|_{T=T_0}$ and $\phi_T^{(s)}=(\partial_T\Phi^{(s)})|_{T=T_0}$ hold.
In order to be able to uncover the coupling between various modes in \eqref{eq: mult-exp} characterized by their multipole indices $s,\ell,m$, the applied initial data will always be a single mode excitation. In addition, they will be either ``static''  or ``purely dynamical''.

Accordingly, for some fixed values of the $s, \ell, m$ indices, a single mode excitation type of initial data  $(\phi^{(s)},\phi_T^{(s)})$ is called {\it static} (ST) if
\begin{equation}\label{eq: ST}
	(ST)\,:
	\begin{cases}
		\phi^{(s)}(R,\vartheta,\varphi)=\phi_\ell{}^{m}(T_0,R)\cdot {}_s{Y_\ell}{}^m(\vartheta,\varphi)\\
		 \phi_T^{(s)}(R,\vartheta,\varphi)\equiv 0\,,
	\end{cases}
\end{equation}
while it will be called {\it purely dynamical} (PD) if
\begin{equation}\label{eq: PD}
(PD)\,:
\begin{cases}
\phi^{(s)}(R,\vartheta,\varphi)\equiv 0, \\
\phi_T^{(s)}(R,\vartheta,\varphi)=(\phi_{T})_\ell{}^{m}(T_0,R)\cdot {}_s{Y_\ell}{}^m(\vartheta,\varphi)
\end{cases}
\end{equation}
hold on the $T=T_0$ initial data surface, and no summation is meant in either \eqref{eq: ST} or in \eqref{eq: PD}. Because of the linearity of the TME \eqref{eq:teurt} and the use of single mode excitations, with multipole indices $s,\ell,m$, all the excited modes in \eqref{eq: mult-exp} will share the $s$ and $m$ values. In what follows, the $\ell$ parameter of the exciting mode will be distinguished by priming it, i.e.,~denoting it by $\ell'$,  while the $\ell$ parameter of the excited modes will be referred to without priming.

The $R$ dependence of the nonvanishing part of the initial data---$\phi_{\ell'}{}^{m}(T_0,R)$ in the  static case or $(\partial_T\phi)_{\ell'}{}^{m}(T_0,R)$ in the purely dynamical case---is restricted
by choosing it to be the ``bump'' with center $c$ and width $w$,
\begin{equation}\label{eq: initial-data-bump}
\mathcal{B}(R) =
\begin{cases}
\frac{2R}{1-R^2}\exp\left(-\frac{1}{R-c+w/2}+\frac{1}{R-c-w/2}+\frac{4}{w}\right)\,,&\mbox{if}\quad c-w/2<R<c+w/2,\\
\hfill 0\,,&\mbox{otherwise\,.}
\end{cases}
\end{equation}
Note that $\mathcal{B}(R)$ is a smooth function of compact support, the parameters of which were fixed in our numerical simulations as $c=0.7$ and $w=0.1$.

\subsection{The late-time behavior}

Likewise, in the case of spin zero fields---when monitored at certain fixed spatial locations---after an initial dynamical phase, each of the excited modes $\phi_\ell{}^{m}(T,R)$ go through a lasting quasinormal ringing period which is supplemented by a late-time tail behavior. In particular, this means that for sufficiently large values of $T$, the individual multipole expansion coefficients at any fixed $R=R_0$ spatial location decrease as
\begin{equation}
\phi_{\ell}{}^{m}(T,R_0) \sim T^{-n}\,,
\end{equation}
where $n$ is a positive integer, the value of which may depend on all the involved parameters; i.e., in general, it has the functional form $n=n(s,\ell,m,\ell')$.

\subsubsection{Earlier analytic and numerical results on the decay rates}
\label{sec: analytic-estimates}

In studying the tail behavior the functional form  of the decay rate $n=n(s,\ell,m,\ell')$ is the center of interest. In interpreting our numerical findings, we may use as our reference frames two independent investigations prior to ours. First, we may refer to the detailed analytic studies carried out by Hod 
\cite{Hod:2000fh}. Second, the accurate numerical investigations carried out in \cite{Harms:2013ib} provide us important clues concerning the functional dependence of the decay rates.

Both of these investigations have some limitations in their scopes. For instance, the decay rates determined in 
\cite{Hod:2000fh} by using some Green's-function-based analytic arguments apply only to purely dynamical initial data; i.e.,~no decay rates are derived therein for static initial data.
It is worth also mentioning that---based on  analogous analytic investigations carried out in \cite{Harms:2013ib} (see, in particular, Appendix B therein)---in certain (though very limited) subcases, these estimates of Hod were claimed to be imprecise. The numerical studies in \cite{Harms:2013ib} were also somewhat restricted, as apart from a few special cases, they were limited to axisymmetric configurations.

Despite the slight limitations in their scopes---as the aforementioned investigations are also complementary to each other---they provide us important guidance in carrying out and interpreting our numerical results.
Below, we summarize the most important findings reported in \cite{Hod:2000fh,Harms:2013ib}.
In particular, for purely dynamical initial data the pertinent results obtained by Green's-function-based analytic arguments
and by numerical simulations can be summarized as follows:
\begin{enumerate}
	\item[(1)] At the horizon $R=R_+$,
	\begin{equation}\label{eq: PD1}
        n|_{R_+} = \begin{cases}
    	\ell'+\ell+3+\alpha\,,& \mathrm{if}\ \ \ell'=\ell_0,\\
        \ell'+\ell+3+\alpha-\delta\,,& \mathrm{if}\ \ \ell'=\ell_0+1,\\
    	\ell'+\ell+1+\alpha\,,&\mathrm{if}\ \ \ell'>\ell_0+1\,,
    	\end{cases}
	\end{equation}
    where $\ell_0=\max\{|m|,|s|\}$ is the lowest allowed value of $\ell$ for given $s$ and $m$, and $\alpha=0$ in all cases except if $s>0$ and $m=0$ when $\alpha=1$. We find $\delta=1$ in the analytic investigations of Hod \cite{Hod:2000fh} but numerical results in \cite{Harms:2013ib} suggest that at least in the case $m=0$, the correct value is $\delta=0$.

	\item[(2)] At finite intermediate spatial locations with $R_+<R<1$, the values of $n$ can be deduced from \eqref{eq: PD1} by the substitution of $\alpha=0$, i.e.,
	\begin{equation}\label{eq: PD2}
        n|_{R} = \begin{cases}
        \ell'+\ell+3\,,& \mathrm{if}\ \ \ell'=\ell_0,\\
        \ell'+\ell+3-\delta\,,& \mathrm{if}\ \ \ell'=\ell_0+1,\\
        \ell'+\ell+1\,,&\mathrm{if}\ \ \ell'>\ell_0+1\,.
        \end{cases}
	\end{equation}
	Similarly, as in \eqref{eq: PD1} $\delta=1$ in \cite{Hod:2000fh}; however, in this case, in Appendix B of \cite{Harms:2013ib} Harms et al. pointed out a missed case in the argument of Hod, and as a result, $\delta=0$ when $m=0$.
	This correction is also consistent with the numerical results of \cite{Harms:2013ib}.

	\item[(3)] At $R=1$, representing future null infinity $\mathscr{I}^+$,
	\begin{equation}\label{eq: PD3}
	\hskip-2.1cm
	n|_{R=1} = \begin{cases}
	\ell-s+2+\gamma\,,& \mathrm{if}\ \ \ell' \le \ell+1\,, \\
	\ell'-s\,,& \mathrm{if}\ \ \ell' > \ell+1 \,,
	\end{cases}
	\end{equation}
	where  $\gamma=0$ in all cases except if $m=0$, $\ell'=\ell_0+1$, and $\ell=\ell_0$ when $\gamma=1$.
    With vanishing $\gamma$, (\ref{eq: PD3}) reproduces the result of Hod \cite{Hod:2000fh}; nevertheless, in  Appendix B of \cite{Harms:2013ib} the authors pointed out a missed case in the corresponding argument of \cite{Hod:2000fh}, necessitating the inclusion of $\gamma$ in (\ref{eq: PD3}).
    This correction was also verified numerically in \cite{Harms:2013ib}.
\end{enumerate}
Although the numerical results presented in \cite{Harms:2013ib} are mostly for axisymmetric configurations,
there are two sets of decay exponents gained from nonaxisymmetric purely dynamical initial data
with $s=0$, $m=1$ and $s=-2$, $m=2$, respectively.
The exponents for $s=0$, $m=1$ and $s=-2$, $m=2$ are in agreement with the predictions of \cite{Hod:1999rx,Hod:2000fh}.

\bigskip

In the case of static initial data, we may only refer to the empirical results in \cite{Harms:2013ib} where, as mentioned above, the functional forms are limited to axially symmetric configurations with $m=0$. The pertinent findings in \cite{Harms:2013ib} can be summarized as follows:

\begin{enumerate}
	\item[(1')] At the horizon $R=R_+$,
	\begin{equation}\label{eq: ST1}
	n|_{R_+} = \begin{cases}
	\ell'+\ell+3+\alpha\,,& \mathrm{if}\ \ \ell'=|s|\,,\\
	\ell'+\ell+2+\alpha\,,&\mathrm{if}\ \ \ell'>|s|\,,
	\end{cases}
	\end{equation}
	where now $\alpha=0$ for $s=0$ and $\alpha=1$ otherwise.
	\item[(2')] At any finite intermediate spatial location with $R_+<R<1$,  the values of $n$ can be given as in \eqref{eq: ST1}, with the distinction that $\alpha=0$ in all possible cases, i.e.,
	\begin{equation}\label{eq: ST2}
	n|_{R} = \begin{cases}
	\ell'+\ell+3\,,& \mathrm{if}\ \ \ell'=|s|\,\\
	\ell'+\ell+2\,,&\mathrm{if}\ \ \ell'>|s|\,.
	\end{cases}
	\end{equation}
	\item[(3')] At $R=1$,
	\begin{equation}\label{eq: ST3}
	\hskip-1.1cm
	n|_{R=1} = \begin{cases}
	\ell-s+2\,,& \mathrm{if}\ \ \ell'\leq \ell\,,\\
	\ell'-s+1\,,&\mathrm{if}\ \ \ell' > \ell\,.
	\end{cases}
	\end{equation}
\end{enumerate}

As both the analytic and numerical setups outlined in the previous sections allow us to study nonaxisymmetric configurations, one of our main motivations in this paper is, besides verifying \eqref{eq: ST1}--\eqref{eq: ST3}, if possible, to deduce the corresponding generalizations of these relations.

\subsubsection{The local power index}
\begin{figure}[!ht]
	\begin{center}
        \includegraphics[width=.6\textwidth]{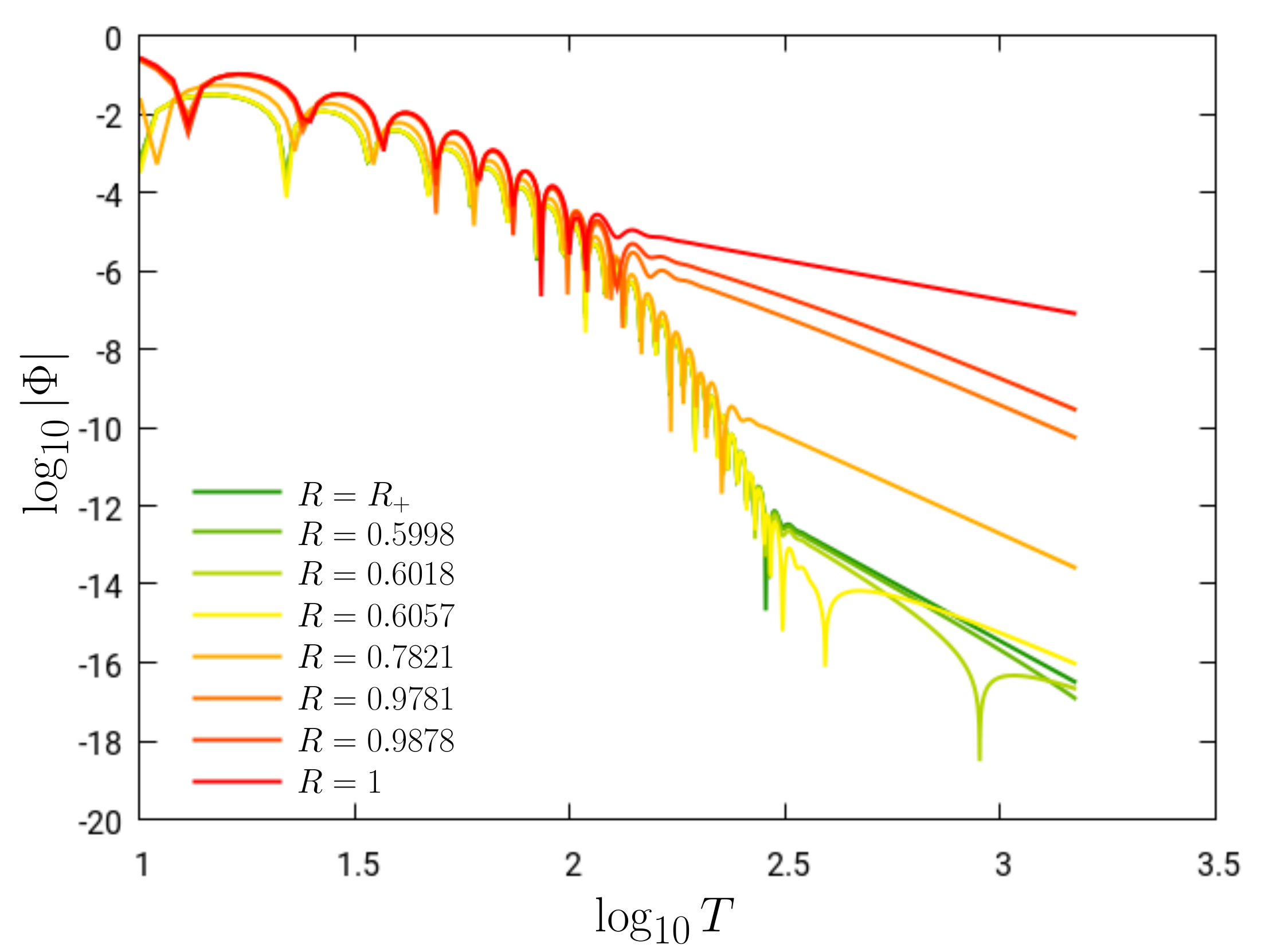}
		\caption{\small
			For the specific choice of excitation with multipole parameters $s=-1, \ell'=1, m=0$, the time dependence of the $\ell=1$ mode is depicted at the outer horizon $R=R_+$ at the $R=1$ line representing future null infinity and at some intermediate $R=const$ locations. At each fixed spatial location, the initial long-lasting quasinormal oscillatory phase is supplemented by a specific power-law decay corresponding to some integer rate $n$.}
		\label{fig:loglog}
	\end{center}
\end{figure}

As mentioned above, the excited modes following a long-lasting quasinormal ringing phase all end up in a power-law decay $\phi_{\ell}{}^{m}\sim T^{-n}$ \cite{Harms:2013ib, Hod:2000fh, Racz:2011qu} as depicted in Fig.~\ref{fig:loglog}.
The question is then how to determine the specific value of $n$. In practice, the value of the decay rate $n$---for each specific mode---is approximated by the local power index (LPI) $\mu$ determined as
\begin{equation}\label{eq: LPI}
	\mu=\frac{\partial\ln |\phi|}{\partial\ln T}=T\cdot\frac{\mathfrak{Re}(\phi)\cdot\mathfrak{Re}(\phi_T)+\mathfrak{Im}(\phi)\cdot\mathfrak{Im}(\phi_T) }{(\mathfrak{Re}(\phi))^2+(\mathfrak{Im}(\phi))^2}\,,
\end{equation}
where, for simplicity, the multipole indices $\ell$ and $m$ of $\phi_{\ell}{}^{m}$ are suppressed.
Note that, just as the decay rate $n$, the local power index $\mu$ also depends on the spatial location as well as on the multipole indices $s$, $\ell, m$, and $\ell'$ of the involved modes \cite{Harms:2013ib, Racz:2011qu}.

\section{Numerical results}\label{sec: numerical-results}

The discussions up to this point are valid for any linear field of spin $s$ satisfying the homogeneous Teukolsky master equation on a Kerr black hole background. In this section, considerations will be restricted to the numerical investigations of linear fields of spin $s=\pm 1,\pm 2$, and our numerical findings will be reviewed.

\subsection{More specific settings}

\subsubsection{The input parameters}

\begin{figure}[!ht]
	\begin{center}
        \includegraphics[width=.6\textwidth]{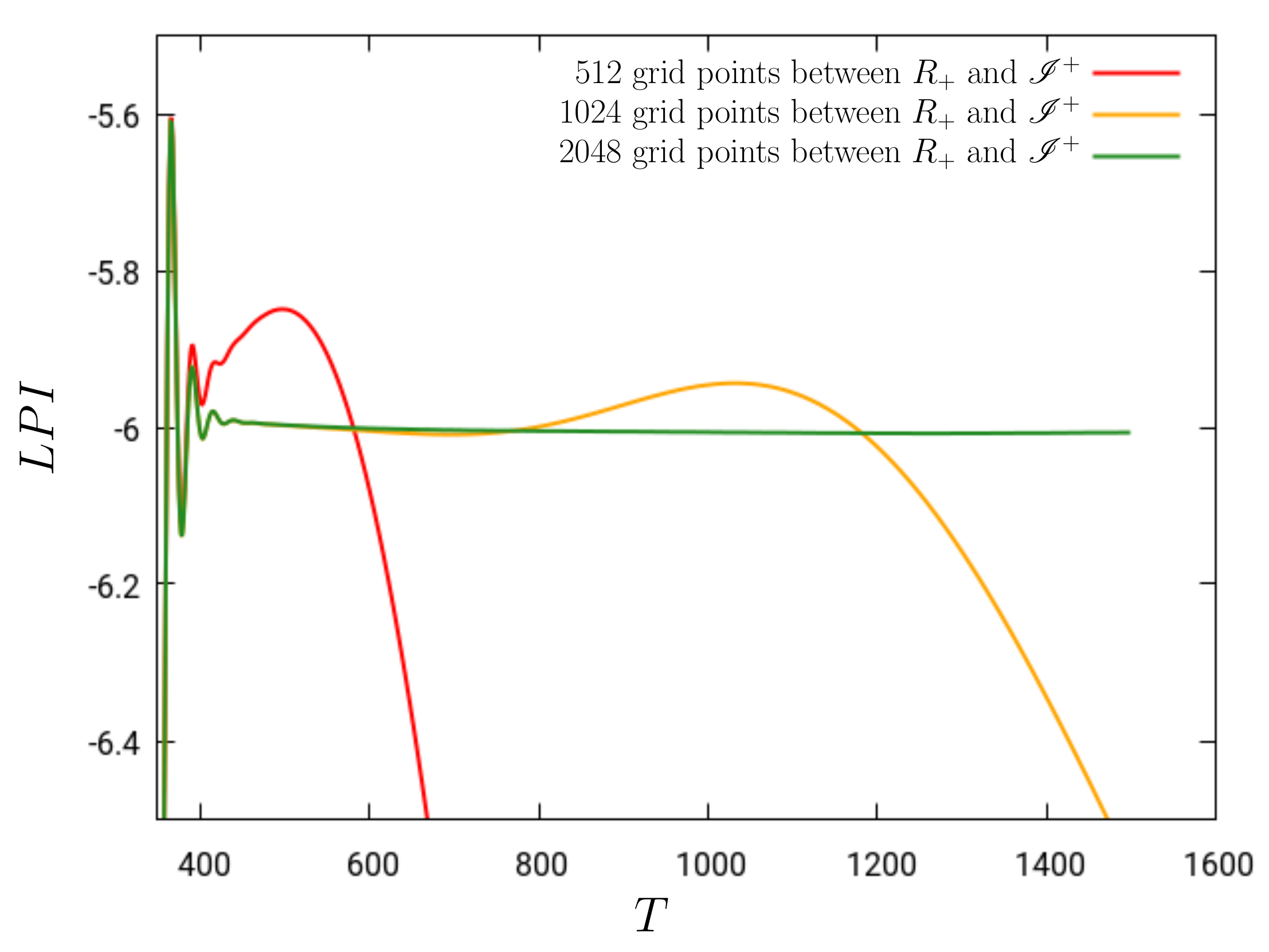}
		\caption{\small The $T$ dependence of the LPI of the $\ell=1$ excited mode is depicted for various resolutions. The multipole indices of applied static excitation were $s=1$, $\ell'=1$, $m=1$. It is visible that the use of $512$ spatial grid points does not even allow us to get a hint of the correct value of LPI. It is also clearly visible that by increasing the number of involved grid points the value of $\mu$ becomes more accurate.}
		\label{fig:lpiconv}
	\end{center}
\end{figure}

In most of our numerical simulations the parameters of the Kerr background were fixed as  $M=1$ and $a=0.5$. A cutoff in the multipole expansion \eqref{eq: mult-exp} at $\ell_{max}=8$ was found to be completely satisfactory for
achieving the necessary precision. Some experimentation with doubling $\ell_{max}$ shows that the accuracy of our results is not limited by this value. This suggests that the error introduced by
radial discretization and numerical arithmetic is still bigger than the error introduced by this cutoff.
In many of the simulations, the use of $1024$ spatial grid points on the $[R_+,1]$ interval was satisfactory (see Fig.\,\ref{fig:lpiconv}), though, in most cases we used $2048$, and there were also cases where the use $4098$ spatial grid points was more rewarding.
As for the applied numerical precision, note that in most of our simulations the use of \emph{long double}  arithmetic was sufficient. Nevertheless, as for negative values of the spin parameter $s$, the LPI values were always harder to be determined, and we used the computationally more expensive \emph{quadruple} precision in those cases. Unfortunately, for negative $s$ values even the use of quadruple precision could not always guarantee the required level of precision at late time which, on the other hand, is essential to draw a conclusive result on the tail behavior. In many cases, the pertinent LPIs could not even be determined despite that for positive values of the spin parameter the LPI values are already sufficiently accurate with the use of long double precision and, in general, with the use of lower spatial resolution.

\subsubsection{Assembling the determined LPIs}

In the succeeding subsections, our numerical findings will be reported. First, the LPIs determined are collected in tables,
and then the implications of the observed phenomena are described briefly.

The tables of LPIs are given for various values of the relevant parameters. In particular, $s=\pm 1, \pm 2$ with exciting modes $\ell'$ and excited modes $\ell$ both taking (the allowed) positive integer values from the interval $1\leq \ell',\ell \leq 5$. The values of $m$ will also be restricted to $0,\pm 1,\pm 2$.

\begin{figure}[!ht]
	\begin{center}
        \includegraphics[width=.6\textwidth]{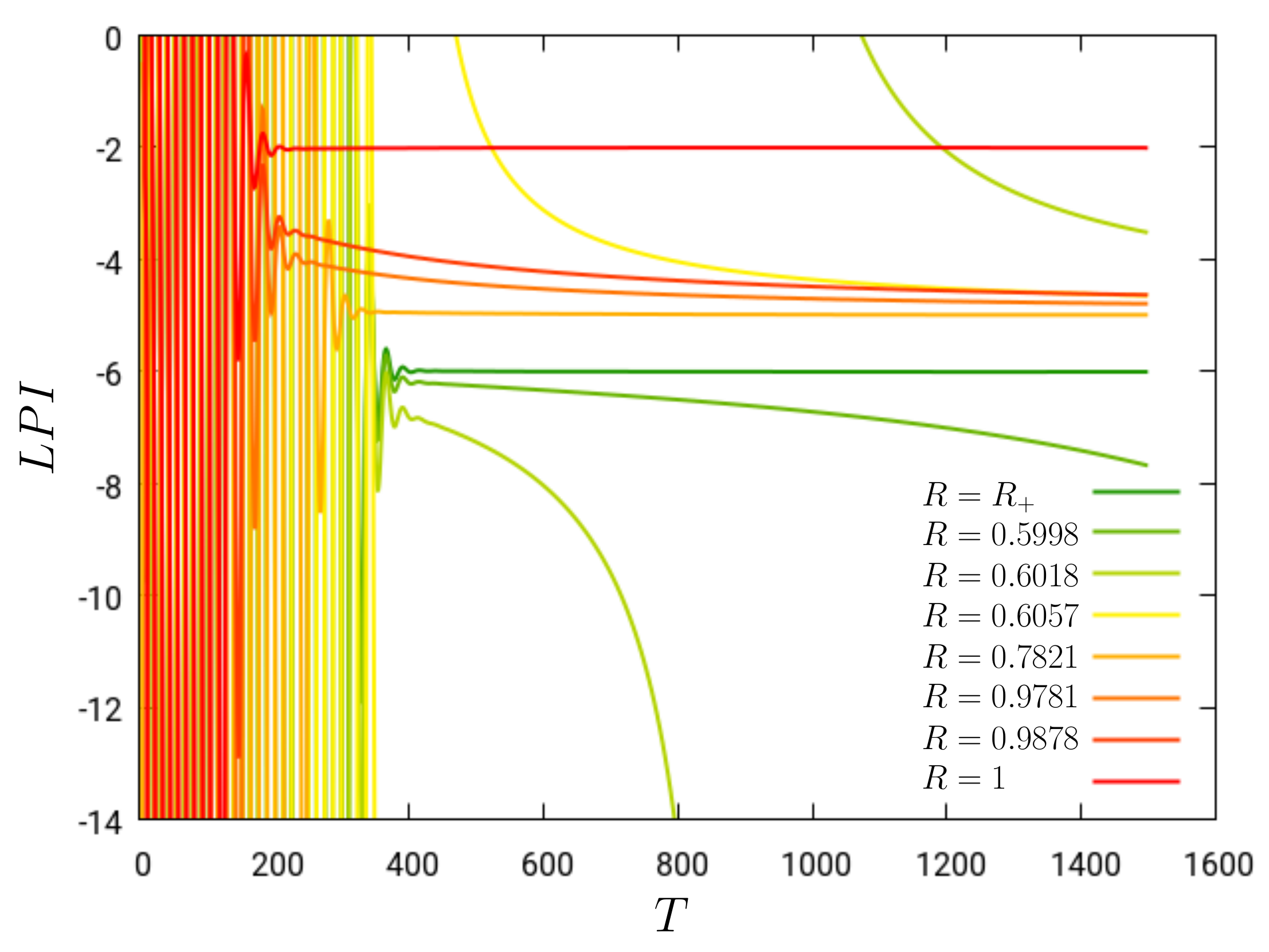}
		\caption{\small
			The $R$ dependence of the LPI values is depicted for a mode with multipole indices $s=1, \ell=1, m=0$ that was yielded by applying a static initial excitation with $\ell'=1$. The value of $\mu$ at the outer horizon $R=R_+$ is $-6$, while it is $-2$ at the $R=1$ line representing future null infinity. It is also clearly indicated that the closer a finite intermediate spatial location with $R_+<R<1$ is to the outer horizon $R=R_+$ or to the $R=1$ line representing future null infinity, the longer it takes to settle down to the pertinent shared LPI value $\mu=-5$.}
		\label{fig:lpiRdep}
	\end{center}
\end{figure}

In advance of turning to the contents of these tables, note first that concerning the dependences of the LPIs on the spin parameter $s$, the following two cases have significantly different characters. Whenever $s<0$, it suffices to determine the values of $\mu$ at the outer horizon $R=R_+$ and at the future null infinity $R=1$, as the LPIs relevant for the intermediate values $R_+<R<1$ are exactly the same as those at the outer horizon $R=R_+$. As opposed to this, whenever $s>0$, the LPIs have a higher variety; i.e.,~the value of $\mu$ at the intermediate location $R_+<R<1$ differs from that at the outer horizon $R=R_+$ though asymptotically the LPI takes the same value for any intermediate finite locations $R_+<R<1$. As it is clearly indicated by the graphs in Fig.\,\ref{fig:lpiRdep}, for $s>0$, the values of  $\mu$ have to be monitored not only at the outer horizon ($R=R_+$) and at future null infinity ($R=1$)
but for several intermediate $R=const$ locations as well.
Remarkably, the closer the intermediate $R$ value is to $R=R_+$ or to $R=1$, the longer it takes to settle down to the LPI value that is relevant for pertinent intermediate locations.

\medskip

In accordance with the above outlined observations, the tables will be structured as follows: As for negative $s$ (with $s=-1$ or $s=-2$), the LPI values at the horizon and intermediate finite locations are the same, and for each slot labeled by $\ell'$ and $\ell$, only the values ``$\mu_{R_+}$'' at the horizon and ``$\mu_{\scri^+}$'' at future null infinity will be indicated by separating them with a vertical line in writing ``$\mu_{R_+}|\,\mu_{\scri^+}$.'' As for the positive $s$ (with $s=1$ or $s=2$), the LPI values at the intermediate locations, with $R_+<R<1$ differ from that of $\mu_{R_+}$ and $\mu_{\scri^+}$, and we arrange these three values by separating them with vertical lines in writing ``$\mu_{R_+}\,|\,\mu_{R}\,|\,\mu_{\scri^+}$.''  In practice, the LPI value indicated in the middle slot will always correspond to the one measured along the $R=0.88$ timeline.

There is various additional information indicated in the succeeding tables. Note first that in most cases the precise value of the LPIs could be determined in a clean way by making use of \eqref{eq: LPI}  (see also Figs.\,\ref{fig:loglog}--\ref{fig:lpiRdep}). Nevertheless, there were some cases where the limited accuracy at the very late time did not allow us to draw a clear conclusion this way. In such cases, instead of using \eqref{eq: LPI}, a line fitting on the log-log plots was applied in order to determine the approximate value of the LPIs relevant for the pertinent modes. Note that, as the output of this method depends, for instance, on the choice of the subintervals where the fitting is carried out, the yielded LPI values have to be taken with some caveats. To warn the reader, the corresponding LPI values are always indicated by round brackets around the pertinent numbers, i.e.,\,by writing $(\mu)$ instead of $\mu$.
We also have to admit that there were cases---especially for negative values of $s$---where neither of the above-discussed methods could be sufficiently conclusive. All these cases will be indicated by filling up
the pertinent slot labeled by the $\ell'$ and $\ell$ values with a question mark ``?.''

\subsection{Axially symmetric configurations}\label{subsec: axially-symm-conf}

Note first that in \cite{Harms:2013ib} detailed numerical investigations of the axisymmetric case were carried out. This gives us the opportunity to compare the observed LPI values and, thereby, to check the performance and the reliability  of our code.

The LPIs obtained for the axially symmetric cases with $|s|=\pm 1,\pm 2$ and with $m=0$ are assembled in Table\,\ref{tab:axisymm-s1} for $|s|=\pm 1$ and in Table\,\ref{tab:axisymm-s2} for $|s|=\pm 2$.

\begin{table}[!ht]
	\centering
	{ 
		\scriptsize
		\begin{subtable}{0.45\textwidth}
			\begin{tabular}{c|ccccc}
				$\ell$'   &$\ell$=1    &2  &3                 &4         &5\\ \hline
				1    &5|4    &6|5          &7|6               &8|7       &9|8\\
				2    &5|4    &6|5          &7|6               &8|(7)     &9|?\\
				3    &6|5    &7|5          &8|(\diffr{5})     & 9|?      &10|?\\
				4    &7|6    &8|6          &9|6               &10|?      &11|?\\
				5    &8|7    &9|7          &10|7              & ?|?      & ?|?
			\end{tabular}
			\caption{$s=-1$, static initial data.}
		\end{subtable}
		\begin{subtable}{0.5\textwidth}
			\begin{tabular}{c|ccccc}
				$\ell$'   &$\ell$=1  &2   &3          &4          &5\\\hline
				1    &6|5|2    & 7|6|3    & 8| 7|4    & 9| 8|5    &10|  9 | 6\\
				2    &6|5|2    & 7|6|3    & 8| 7|4    & 9| 8|5    &10|  9 | 6\\
				3    &7|6|3    & 8|7|3    & 9| 8|4    &10| 9|5    &11| 10 |(\diffr{7})\\
				4    &8|7|4    & 9|8|4    &10| 9|4    &11|10|?    &12| 11 |(\diffr{7})\\
				5    &9|8|5    &10|9|5    &11|10|5    &12|11|5    &13|(12)|(\diffr{7})
			\end{tabular}
			\caption{$s=1$, static initial data.}
		\end{subtable}
		\begin{subtable}{0.45\textwidth}
			\begin{tabular}{c|ccccc}
				$\ell$' &$\ell$=1 &2 &3                &4          &5\\\hline
				1    &5|4    &6|5    &7| 6             &  8 | 7    & 9|(\diffr{9})\\
				2    &6|5    &7|5    &8| 6             &  9 | 7    &10|(8)\\
				3    &5|4    &6|5    &7| 6             &  8 | 7    & 9| ?\\
				4    &6|5    &7|5    &8|(\diffr{5})     &  9 |(7)   &10| ?\\
				5    &7|6    &8|6    &9| 6             &(10)| ?    & ?| ?
			\end{tabular}
			\caption{$s=-1$, dynamic initial data.}
		\end{subtable}
		\begin{subtable}{0.5\textwidth}
			\begin{tabular}{c|ccccc}
				$\ell$' &$\ell$=1 &2     &3         &4                    &5\\\hline
				1    &6|5|2    &7|6|3    & 8|7|4    & 9| 8| 5             &10|  9 |6\\
				2    &7|6|3    &8|7|3    & 9|8|4    &10| 9| 5             &11| 10 |6\\
				3    &6|5|2    &7|6|3    & 8|7|4    & 9| 8| 5             &10| (8)|6\\
				4    &7|6|3    &8|7|3    & 9|8|4    &10| 9| 5             &11| 10 |6\\
				5    &8|7|4    &9|8|4    &10|9|4    &11|10|(\diffr{4})     &12|(10)|6
			\end{tabular}
			\caption{$s=1$, dynamic initial data.}
		\end{subtable}
	}
	\caption{$|s|=1$, $m=0$
	}
	\label{tab:axisymm-s1}
\end{table}
\begin{table}
	\centering
	{\scriptsize
		\begin{subtable}{0.45\textwidth}
			\begin{tabular}{c|ccccc}
				$\ell$' &$\ell$=1 &2        &3        &4         &5\\ \hline
				1    &X    &X               &X        &X         &X\\
				2    &X    &7|6             &8|?      &9|?       &10|?\\
				3    &X    &7|6             &8|?      &9|?       &10|?\\
				4    &X    &8|7             &?|?      &?|?       & ?|?\\
				5    &X    &9|8             &?|?      &?|?       & ?|?
			\end{tabular}
			\caption{$s=-2$, static initial data.}
		\end{subtable}
		\begin{subtable}{0.5\textwidth}
			\begin{tabular}{c|ccccc}
				$\ell$' &$\ell$=1 &2  &3          &4          &5\\\hline
				1    &X    & X        & X         & X         &X\\
				2    &X    & 8|7|2    & 9| 8|3    &10| 9|4    &11|10|5\\
				3    &X    & 8|7|2    & 9| 8|3    &10| 9|4    &11|10|5\\
				4    &X    & 9|8|3    &10| 9|3    &11|10|4    &12|11|5\\
				5    &X    &10|9|4    &11|10|4    &12|11|4    &13|12|5
			\end{tabular}
			\caption{$s=2$, static initial data.}
		\end{subtable}
		\begin{subtable}{0.45\textwidth}
			\begin{tabular}{c|ccccc}
				$\ell$' &$\ell$=1 &2        &3        &4         &5\\ \hline
				1    &X    &X               &X        &X         &X\\
				2    &X    &7|6             &8|7      &9|8       &10|9\\
				3    &X    &8|7             &?|?      &?|?       & ?|?\\
				4    &X    &7|6             &8|?      &9|?       &10|?\\
				5    &X    &8|7             &?|?      &?|?       & ?|?
			\end{tabular}
			\caption{$s=-2$, dynamic initial data.}
		\end{subtable}
		\begin{subtable}{0.5\textwidth}
			\begin{tabular}{c|ccccc}
				$\ell$' &$\ell$=1 &2     &3         &4          &5\\\hline
				1    &  X      &  X      &  X       &   X       &    X  \\
				2    &  X      &8|7|2    & 9|8|3    &10| 9|4    &11|10|5\\
				3    &  X      &9|8|3    &10|9|3    &11|10|4    &12|11|5\\
				4    &  X      &8|7|2    & 9|8|?    &10| 9|4    &11|10|5\\
				5    &  X      &9|8|3    &10|9|3    &11|10|4    &12|11|5
			\end{tabular}
			\caption{$s=2$, dynamic initial data.}
		\end{subtable}
	}
	\caption{$|s|=2$, $m=0$}
	\label{tab:axisymm-s2}
\end{table}

The most characteristic features which deserve notice are as follows:
\begin{enumerate}
 	\item The values of the LPIs at $R_+$ and at intermediate locations $R_+<R<1$ are always increased by $1$ if the value of the excited mode $\ell$ is increased by $1$.
 	\item The values of the LPIs at $\scri^+$ do not necessarily increase in an analogous, strictly monotonous way, though they never decrease either while the value of $\ell$ is increased.
	\item The above two observations imply that the mode with the lowest possible $\ell$ value always decays at the slowest rate, though at $\scri^+$ there may be some other $\ell$ modes which are also decaying at the same rate.
	\item For $s>0$, the value $\mu_{R_+}$ is always larger by $1$ than the LPI at intermediate locations $R_+<R<1$.
	\item All observations made in the previous point are in full agreement with the analytic results and numerical findings in \cite{Hod:2000fh} and \cite{Harms:2013ib}, respectively.
	\item For $|s|=2$, the most significant distinction is that the first row and the first column of the tables signified by $\ell'$, $\ell$ slots are missing, as the corresponding modes are excluded by the fact that all the spin-weighted spherical harmonics with $\ell<\max\{|s|,|m|\}$ are annihilated by the $\eth$ operator.
\end{enumerate}

Note, finally, that in Table\,\ref{tab:axisymm-s1} there are seven numerically determined LPI values---they are indicated by boldface (in color by orange)---that differ from those which can be deduced from the rules laid down in \cite{Hod:2000fh,Harms:2013ib}. Notably, the difference is always $1$. Nevertheless, these seven numbers are all in round brackets indicating that the corresponding estimates have to be taken with some caveats.

 \subsection{Nonaxially symmetric configurations}\label{subsec: non-axially-symm-conf}

Note that in the nonaxisymmetric scenarios, due to the fact that the spin-weighted spherical harmonics with $\ell<\max\{|s|,|m|\}$ are annihilated by the $\eth$ operator the first row and the first column of the tables signified by the $\ell'$, $\ell$ slots are always missing if either $|m|=2$ or $|s|=2$. Nevertheless, for the sake of easier comparisons between the axisymmetric and nonaxisymmetric cases, we assemble the LPI values using the interval $1\leq \ell',\ell \leq 5$, as done previously.

\subsubsection{$|m|=1,2$ with static initial data}

In this subsection, excitations generated by the $|m|=1,2$ modes with static initial data are considered. The relevant LPIs are collected in Tables \ref{tab:m1stat}--\ref{tab:s2m2stat}, and the main observations concerning them are summarized as follows:
\begin{table}[!ht]
 \centering
 {\scriptsize
  \begin{subtable}{0.45\textwidth}
  \begin{tabular}{c|ccccc}
   $\ell$'   &$\ell$=1    &2         &3      &4         &5\\\hline
   1    &5|4    &6|5       &7|6    &8|7       &9|?\\
   2    &5|4    &6|5       &7|6    &8|(7)     &9|?\\
   3    &6|5    &7|5       &8|6    &9|(\diffr{6})   &10|?\\
   4    &7|6    &8|6       &9|6    &10|?       &11|?\\
   5    &8|7    &9|7      &10|7    &?|?      & ?|?
  \end{tabular}
  \caption{$s=-1$, $m=1$}
  \label{tab:01}
  \end{subtable}
  \begin{subtable}{0.5\textwidth}
  \begin{tabular}{c|ccccc}
   $\ell$'   &$\ell$=1      &2         &3          &4          &5\\\hline
   1    &5|5|2    & 6|6|3    & 7| 7|4    & 8| 8|5    & 9|  9 | 6\\
   2    &5|5|2    & 6|6|3    & 7| 7|4    & 8| 8|5    & 9|  9 | 6\\
   3    &6|6|3    & 7|7|3    & 8| 8|4    & 9| 9|5    &10| 10 |6\\
   4    &7|7|4    & 8|8|4    & 9| 9|4    &10|10|5    &11| 11 |(6)\\
   5    &8|8|5    & 9|9|5    &10|10|5    &11|11|5    &12|(12)|(6)
  \end{tabular}
  \caption{$s=1$, $m=1$}
  \end{subtable}
  \begin{subtable}{0.45\textwidth}
  \begin{tabular}{c|ccccc}
   $\ell$'   &$\ell$=1    &2      &3      &4     &5\\\hline
   1    &5|4    &6|5    &7|6    &8|7   &9|?\\
   2    &5|4    &6|5    &7|6    &8|?   &9|?\\
   3    &6|5    &7|5    &8|6    &9|?   &10|?\\
   4    &7|6    &8|6    &9|6    &10|?   &11|?\\
   5    &8|7    &9|7    &10|7   &?|?   &?|?
  \end{tabular}
  \caption{$s=-1$, $m=-1$}
  \end{subtable}
  \begin{subtable}{0.5\textwidth}
  \begin{tabular}{c|ccccc}
   $\ell$'   &$\ell$=1      &2        &3        &4        &5\\\hline
   1    &5|5|2    &6|6|3    &7|7|4    &8|8|5    &9|9|6\\
   2    &5|5|2    &6|6|3    &7|7|4    &8|8|5    &9|9|6\\
   3    &6|6|3    &7|7|3    &8|8|4    &9|9|5    &10|10|6\\
   4    &7|7|4    &8|8|4    &9|9|4    &10|10|5  &11|11|(6)\\
   5    &8|8|5    &9|9|5    &10|10|5  &11|11|5  &12|(12)|(6)
  \end{tabular}
  \caption{$s=1$, $m=-1$}
  \end{subtable}
  }
  \caption{$|s|=1$, $|m|=1$ with static initial data}
  \label{tab:m1stat}
  \end{table}
\begin{table}[!ht]
	\centering
	{\scriptsize
		\begin{subtable}{0.45\textwidth}
			\begin{tabular}{c|ccccc}
				$\ell$'   &$\ell$=1    &2         &3      &4         &5\\\hline
				1    &X    &X         &X      &X       &X\\
				2    &X    &7|5       &8|6    &9|7     &10|?\\
				3    &X    &7|5       &8|6    &9|(7)   &10|?\\
				4    &X    &8|6       &9|6    &10|?    &10|?\\
				5    &X    &9|7       &10|7   &?|?      & ?|?
			\end{tabular}
			\caption{$s=-1$, $m=2$}
		\end{subtable}
		\begin{subtable}{0.5\textwidth}
			\begin{tabular}{c|ccccc}
				$\ell$'   &$\ell$=1      &2         &3          &4          &5\\\hline
				1    &X    & X        & X         & X         & X\\
				2    &X    & 7|7|3    & 8| 8|4    & 9| 9|5    &10| 10 | 6\\
				3    &X    & 7|7|3    & 8| 8|4    & 9| 9|5    &10| 10 | 6\\
				4    &X    & 8|8|4    & 9| 9|4    &10|10|5    &11| 11 |6\\
				5    &X    & 9|9|5    &10|10|5    &11|11|5    &12|(12)|6
			\end{tabular}
			\caption{$s=1$, $m=2$}
		\end{subtable}
		\begin{subtable}{0.45\textwidth}
			\begin{tabular}{c|ccccc}
				$\ell$'   &$\ell$=1    &2      &3      &4     &5\\\hline
				1    &X    &X      &X      &X      &X\\
				2    &X    &7|5    &8|6    &9|7    &10|?\\
				3    &X    &7|5    &8|6    &9|(7)  &10|?\\
				4    &X    &8|6    &9|6    &10|?   &11|?\\
				5    &X    &9|7    &10|7   &?|?    &?|?
			\end{tabular}
			\caption{$s=-1$, $m=-2$}
		\end{subtable}
		\begin{subtable}{0.5\textwidth}
			\begin{tabular}{c|ccccc}
				$\ell$'   &$\ell$=1      &2        &3        &4        &5\\\hline
				1    &X    &X        &X        &X        &X\\
				2    &X    &7|7|3    &8|8|4    &9|9|5    &10|10|6\\
				3    &X    &7|7|3    &8|8|4    &9|9|5  &10|10|6\\
				4    &X    &8|8|4    &9|9|4  &10|10|5  &11|11|6\\
				5    &X    &9|9|5    &10|10|5  &11|11|5  &12|(12)|6
			\end{tabular}
			\caption{$s=1$, $m=-2$}
		\end{subtable}
	}
	\caption{ $|s|=1$, $|m|=2$ with static initial data}
	\label{tab:m2stat}
\end{table}
\begin{table}[!ht]
	\centering
	{\scriptsize
		\begin{subtable}{0.45\textwidth}
			\begin{tabular}{c|ccccc}
				$\ell$' &$\ell$=1 &2    &3      &4            &5\\\hline
				1    & X     & X        & X     & X           & X \\
				2    & X     &7|6       &8|7    &9|8          &10|?\\
				3    & X     &7|6       &8|?    &9|?          &10|?\\
				4    & X     &8|7       &?|?    &?|?          & ?|?\\
				5    & X     &9|8       &?|?    &?|?          & ?|?
			\end{tabular}
			\caption{$s=-2$, $m=1$}
		\end{subtable}
		\begin{subtable}{0.5\textwidth}
			\begin{tabular}{c|ccccc}
				$\ell$' &$\ell$=1 &2      &3          &4          &5\\\hline
				1    &  X      &   X      &   X       &   X       &     X    \\
				2    &  X      & 7|7|2    & 8| 8|3    & 9| 9|4    &10|10|5\\
				3    &  X      & 7|7|2    & 8| 8|3    & 9| 9|4    &10|10|5\\
				4    &  X      & 8|8|3    & 9| 9|3    &10|10|4    &11|11|5\\
				5    &  X      & 9|9|4    &10|10|4    &11|11|4    &12|12|5
			\end{tabular}
			\caption{$s=2$, $m=1$}
		\end{subtable}
		\begin{subtable}{0.45\textwidth}
			\begin{tabular}{c|ccccc}
				$\ell$' &$\ell$=1 &2    &3      &4            &5\\\hline
				1    & X     & X        & X     & X           & X \\
				2    & X     &7|6       &8|7    &9|8          &10|?\\
				3    & X     &7|6       &8|?    &9|?          &10|?\\
				4    & X     &8|7       &?|?    &?|?          & ?|?\\
				5    & X     &9|8       &?|?    &?|?          & ?|?
			\end{tabular}
			\caption{$s=-2$, $m=-1$}
		\end{subtable}
		\begin{subtable}{0.5\textwidth}
			\begin{tabular}{c|ccccc}
				$\ell$' &$\ell$=1 &2     &3          &4          &5\\\hline
				1    &  X      &  X      &  X        &  X        &    X  \\
				2    &  X      &7|7|2    & 8| 8|3    & 9| 9|4    &10|10|5\\
				3    &  X      &7|7|2    & 8| 8|3    & 9| 9|4    &10|10|5\\
				4    &  X      &8|8|3    & 9| 9|3    &10|10|4    &11|11|5\\
				5    &  X      &9|9|4    &10|10|4    &11|11|4    &12|12|5
			\end{tabular}
			\caption{$s=2$, $m=-1$}
		\end{subtable}
	}
	\caption{$|s|=2$, $|m|=1$, static initial data}
	\label{tab:s2m1stat}
\end{table}
\begin{table}[!ht]
	\centering
	{\scriptsize
		\begin{subtable}{0.45\textwidth}
			\begin{tabular}{c|ccccc}
				$\ell$' &$\ell$=1 &2    &3      &4            &5\\\hline
				1    & X     & X        & X     & X           & X \\
				2    & X     &7|6       &8|7    &9|8          &10|9\\
				3    & X     &7|6       &8|?    &9|?          &10|?\\
				4    & X     &8|7       &?|?    &?|?          &?|?\\
				5    & X     &9|8       &?|?    &?|?          &?|?
			\end{tabular}
			\caption{$s=-2$, $m=2$}
		\end{subtable}
		\begin{subtable}{0.5\textwidth}
			\begin{tabular}{c|ccccc}
				$\ell$' &$\ell$=1 &2    &3      &4            &5\\\hline
				1    & X     & X        & X     & X           & X \\
				2    & X     &7|7|2     &8|8|3  &9|9|4        &10|10|5\\
				3    & X     &7|7|2     &8|8|3  &9|9|4        &10|10|5\\
				4    & X     &8|8|3     &9|9|3  &10|10|4      &11|11|5\\
				5    & X     &9|9|4     &10|10|4&11|11|4      &12|12|5
			\end{tabular}
			\caption{$s=2$, $m=2$}
		\end{subtable}
		\begin{subtable}{0.45\textwidth}
			\begin{tabular}{c|ccccc}
				$\ell$' &$\ell$=1 &2    &3      &4            &5\\\hline
				1    & X     & X        & X     & X           & X \\
				2    & X     &7|6       &8|7    &9|8          &10|9\\
				3    & X     &7|6       &8|?    &9|?          &10|?\\
				4    & X     &8|7       &?|?    &?|?          &?|?\\
				5    & X     &9|8       &?|?    &?|?          &?|?
			\end{tabular}
			\caption{$s=-2$, $m=-2$}
		\end{subtable}
		\begin{subtable}{0.5\textwidth}
			\begin{tabular}{c|ccccc}
				$\ell$' &$\ell$=1 &2    &3      &4            &5\\\hline
				1    & X     & X        & X     & X           & X \\
				2    & X     &7|7|2     &8|8|3  &9|9|4        &10|10|5\\
				3    & X     &7|7|2     &8|8|3  &9|9|4        &10|10|5\\
				4    & X     &8|8|3     &9|9|3  &10|10|4      &11|11|5\\
				5    & X     &9|9|4     &10|10|4&11|11|4      &12|12|5
			\end{tabular}
			\caption{$s=2$, $m=-2$}
		\end{subtable}
	}
	\caption{$|s|=2$, $|m|=2$, static initial data}
	\label{tab:s2m2stat}
\end{table}

\begin{enumerate}
	\item Now, even for positive values of $s$, i.e.\,for $s=1$  and $s=2$, the pertinent LPI values at the horizon and the LPI values at intermediate locations coincide.
	\item For any fixed value of  $s$, the LPIs are independent of the sign of $m$.
	\item The LPI values $\mu_{R_+}$ at the horizon are also independent of the sign of  $s$.
	As opposed to this, the LPI values $\mu_{\scri^+}$ at future null infinity depend on the sign of  $s$, such that the values of $\mu_{\scri^+}$ relevant for negative $s$ are always larger by $2\,|s|$ than the pertinent LPI values  $\mu_{\scri^+}$ for positive $s$.
	\item The LPI values  $\mu_{R_+}$ and $\mu_{\scri^+}$ relevant for the considered nonaxisymmetric case are pairwise equal to the  LPI values at intermediate locations $R_+ < R <1$ and $\mu_{\scri^+}$ relevant for the corresponding axially symmetric configurations (discussed in the previous section).
\end{enumerate}

\subsubsection{$|m|=1,2$ with purely dynamical initial data}

The excitations generated by purely dynamical initial data are visibly more interesting. This gets to be transparent, for instance, when $\ell'=\ell_0+1$. In this case the Green's-function-based argument of Hod \cite{Hod:2000fh} was corrected by results in \cite{Harms:2013ib}, though only for $m=0$.
Nevertheless, our findings indicate that corrections are needed also for $m\ne 0$.
The relevant LPI values are highlighted by boldface (in color: green) characters in the Tables \ref{tab:m1dyn}--\ref{tab:s2m2dyn}.
\begin{table}[!ht]
	\centering
	{\scriptsize
		\begin{subtable}{0.45\textwidth}
			\begin{tabular}{c|ccccc}
				$\ell$'   &$\ell$=1     &2                 &3              &4              &5\\\hline
				1    &5|4               &6|5               &7|6            &8|7            &9|(8)\\
				2    &5|4               &6|5               &7|6            &8|7            &?|(8)\\
				3    &5|4               &6|5               &7|6            &8|(7)          &9|?\\
				4    &6|5               &7|5               &8|6            &9|(7)          &10|?\\
				5    &7|6               &8|6               &9|6            &(\diffr{9})|?   & ?|?
			\end{tabular}
			\caption{$s=-1$, $m=1$}
		\end{subtable}
		\begin{subtable}{0.5\textwidth}
			\begin{tabular}{c|ccccc}
				$\ell$' &$\ell$=1 &2      &3          &4          &5\\\hline
				1    &5|5|2    & 6|6|3    & 7| 7|4    & 8| 8|5    & 9|  9 | 6\\
				2    &5|5|2    & 6|\diffg{7}|3        & 7|\diffg{8}|4    & 8|\diffg{9}|5    & 9|\diffg{10}| 6\\
				3    &5|5|2    & 6|6|3    & 7|(7)|4   & 8| ?|5    & 9|  ? |6\\
				4    &6|6|3    & 7|7|3    & 8| 8|4    & 9|9|5     &10| 10 |6\\
				5    &7|7|4    & 8|8|4    & 9| 9|4    &10|10|(5)  &11| 11 |(6)
			\end{tabular}
			\caption{$s=1$, $m=1$}
		\end{subtable}
		\begin{subtable}{0.45\textwidth}
			\begin{tabular}{c|ccccc}
				$\ell$'   &$\ell$=1     &2          &3          &4          &5\\\hline
				1    &5|4               &6|5        &7|6        &8|7        &9|(8)\\
				2    &5|4               &6|5        &7|6        &8|7        &9|(8)\\
				3    &5|4               &6|5        &7|6        &8|(7)      &9|?\\
				4    &6|5               &7|5        &8|6        &9|?        &10|?\\
				5    &7|6               &8|6        &9|6        &10|?       &?|?
			\end{tabular}
			\caption{$s=-1$, $m=-1$}
		\end{subtable}
		\begin{subtable}{0.5\textwidth}
			\begin{tabular}{c|ccccc}
				$\ell$'   &$\ell$=1      &2        &3        &4        &5\\\hline
				1    &5|5|2    &6|6|3    &7|7|4    &8|8|5    &9|9|6\\
				2    &5|5|2    &6|\diffg{7}|3    &7|\diffg{8}|4    &8|\diffg{9}|5    &9|\diffg{10}|6\\
				3    &5|5|2    &6|6|3    &7|?|4    &8|?|5    &9|?|6\\
				4    &6|6|3    &7|7|3    &8|8|4    &9|9|5  &10|10|6\\
				5    &7|7|4    &8|8|4    &9|9|4  &10|10|5  &11|11|6
			\end{tabular}
			\caption{$s=1$, $m=-1$}
		\end{subtable}
	}
	\caption{$|s|=1$, $|m|=1$ with purely dynamical initial data}
	\label{tab:m1dyn}
\end{table}
\begin{table}[!ht]
	\centering
	{\scriptsize
		\begin{subtable}{0.45\textwidth}
			\begin{tabular}{c|ccccc}
				$\ell$'   &$\ell$=1    &2         &3      &4         &5\\\hline
				1    &X    &X         &X      &X       &X\\
				2    &X    &7|5       &8|6    &9|7     &10|?\\
				3    &X    &7|5       &\diffg{9}|6    &\diffg{10}|7    &\diffg{11}|?\\
				4    &X    &7|5       &8|6    &9|?     &10|?\\
				5    &X    &8|6       &9|6    &10|?    &11|?
			\end{tabular}
			\caption{$s=-1$, $m=2$}
		\end{subtable}
		\begin{subtable}{0.5\textwidth}
			\begin{tabular}{c|ccccc}
				$\ell$'   &$\ell$=1      &2         &3          &4          &5\\\hline
				1    &X    & X        & X         & X         & X\\
				2    &X    & 7|7|3    & 8| 8|4    & 9| 9|5    &10| 10 | 6\\
				3    &X    & 7|7|3    & 8|\diffg{ 9}|4    & 9|\diffg{10}|5    &10|\diffg{11} | 6\\
				4    &X    & 7|7|3    & 8| 8|4    & 9| ?|5    &10|  ? |6\\
				5    &X    & 8|8|4    & 9| 9|4    &10|10|5    &11|(11)|6
			\end{tabular}
			\caption{$s=1$, $m=2$}
		\end{subtable}
		\begin{subtable}{0.45\textwidth}
			\begin{tabular}{c|ccccc}
				$\ell$'   &$\ell$=1    &2      &3      &4     &5\\\hline
				1    &X    &X      &X      &X     &X\\
				2    &X    &7|5    &8|6    &9|7   &10|?\\
				3    &X    &7|5    &?|6    &?|7   &?|?\\
				4    &X    &7|5    &8|6    &9|?   &10|?\\
				5    &X    &8|6    &9|6   &10|?   &11|?
			\end{tabular}
			\caption{$s=-1$, $m=-2$}
		\end{subtable}
		\begin{subtable}{0.5\textwidth}
			\begin{tabular}{c|ccccc}
				$\ell$'   &$\ell$=1      &2        &3        &4        &5\\\hline
				1    &X    &X        &X        &X        &X\\
				2    &X    &7|7|3    &8|8|4    &9|9|5    &10|10|6\\
				3    &X    &7|7|3    &8|\diffg{9}|4    &9|\diffg{10}|5  &10|\diffg{11}|6\\
				4    &X    &7|7|3    &8|8|4    &9| ?|5  &10| ?|6\\
				5    &X    &8|8|4    &9|9|4    &10|10|5  &11|?|6
			\end{tabular}
			\caption{$s=1$, $m=-2$}
		\end{subtable}
	}
	\caption{$|s|=1$, $|m|=2$ with purely dynamical initial data}
	\label{tab:m2dyn}
\end{table}
\begin{table}[!ht]
	\centering
	{\scriptsize
		\begin{subtable}{0.45\textwidth}
			\begin{tabular}{c|ccccc}
				$\ell$' &$\ell$=1 &2    &3      &4            &5\\\hline
				1    & X     & X        & X     & X           & X \\
				2    & X     &7|6       &8|7    &9|8          &10|9\\
				3    & X     &?|?       &?|?    &?|?          &?|?\\
				4    & X     &7|6       &8|?    &9|?          &10|?\\
				5    & X     &8|7       &?|?    &?|?          &?|?
			\end{tabular}
			\caption{$s=-2$, $m=1$}
		\end{subtable}
		\begin{subtable}{0.5\textwidth}
			\begin{tabular}{c|ccccc}
				$\ell$' &$\ell$=1 &2      &3          &4          &5\\\hline
				1    &  X      &   X      &   X       &   X       &     X    \\
				2    &  X      & 7|7|2    & 8| 8|3    & 9| 9|4    &10|10|5\\
				3    &  X      & 7|7|2    & 8|\diffg{ 9}|3    & 9|\diffg{10}|4    &10|\diffg{11}|5\\
				4    &  X      & 7|7|2    & 8| 8|3    & 9| 9|4    &10|10|5\\
				5    &  X      & 8|8|3    & 9| 9|3    &10|10|4    &11|11|5
			\end{tabular}
			\caption{$s=2$, $m=1$}
		\end{subtable}
		\begin{subtable}{0.45\textwidth}
			\begin{tabular}{c|ccccc}
				$\ell$' &$\ell$=1 &2    &3      &4            &5\\\hline
				1    & X     & X        & X     & X           & X \\
				2    & X     &7|6       &8|7    &9|8          &10|9\\
				3    & X     &?|?       &?|?    &?|?          &?|?\\
				4    & X     &7|6       &8|?    &9|?          &10|?\\
				5    & X     &8|7       &?|?    &?|?          &?|?
			\end{tabular}
			\caption{$s=-2$, $m=-1$}
		\end{subtable}
		\begin{subtable}{0.5\textwidth}
			\begin{tabular}{c|ccccc}
				$\ell$' &$\ell$=1 &2     &3          &4          &5\\\hline
				1    &  X      &  X      &  X        &  X        &    X  \\
				2    &  X      &7|7|2    & 8| 8|3    & 9| 9|4    &10|10|5\\
				3    &  X      &7|7|2    & 8|\diffg{ 9}|3    & 9|\diffg{10}|4    &10|\diffg{11}|5\\
				4    &  X      &7|7|2    & 8| 8|3    & 9| 9|4    &10|10|5\\
				5    &  X      &8|8|3    & 9| 9|3    &10|10|4    &11|11|5
			\end{tabular}
			\caption{$s=2$, $m=-1$}
		\end{subtable}
	}
	\caption{$|s|=2$, $|m|=1$, purely dynamical initial data}
	\label{tab:s2m1dyn}
\end{table}
\begin{table}[!ht]
	\centering
	{\scriptsize
		\begin{subtable}{0.45\textwidth}
			\begin{tabular}{c|ccccc}
				$\ell$' &$\ell$=1 &2    &3      &4            &5\\\hline
				1    & X     & X        & X     & X           & X \\
				2    & X     &7|6       &8|7    &9|8          &10|9\\
				3    & X     &7|6       &?|?    &?|?          &?|?\\
				4    & X     &7|6       &8|?    &9|?          &10|?\\
				5    & X     &8|7       &?|?    &?|?          &?|?
			\end{tabular}
			\caption{$s=-2$, $m=2$}
		\end{subtable}
		\begin{subtable}{0.5\textwidth}
			\begin{tabular}{c|ccccc}
				$\ell$' &$\ell$=1 &2    &3      &4            &5\\\hline
				1    & X     & X        & X     & X           & X \\
				2    & X     &7|7|2     &8|8|3  &9|9|4        &10|10|5\\
				3    & X     &7|7|2     &8|\diffg{9}|3  &9|\diffg{10}|4       &10|\diffg{11}|5\\
				4    & X     &7|7|2     &8|8|3  &9|9|4        &10|10|5\\
				5    & X     &8|8|3     &9|9|3  &10|10|4      &11|11|5
			\end{tabular}
			\caption{$s=2$, $m=2$}
		\end{subtable}
		\begin{subtable}{0.45\textwidth}
			\begin{tabular}{c|ccccc}
				$\ell$' &$\ell$=1 &2    &3      &4            &5\\\hline
				1    & X     & X        & X     & X           & X \\
				2    & X     &7|6       &8|7    &9|8          &10|9\\
				3    & X     &7|6       &?|?    &?|?          &?|?\\
				4    & X     &7|6       &8|?    &9|?          &10|?\\
				5    & X     &8|7       &?|?    &?|?          &?|?
			\end{tabular}
			\caption{$s=-2$, $m=-2$}
		\end{subtable}
		\begin{subtable}{0.5\textwidth}
			\begin{tabular}{c|ccccc}
				$\ell$' &$\ell$=1 &2    &3      &4            &5\\\hline
				1    & X     & X        & X     & X           & X \\
				2    & X     &7|7|2     &8|8|3  &9|9|4        &10|10|5\\
				3    & X     &7|7|2     &8|\diffg{9}|3  &9|\diffg{10}|4       &10|\diffg{11}|5\\
				4    & X     &7|7|2     &8|8|3  &9|9|4        &10|10|5\\
				5    & X     &8|8|3     &9|9|3  &10|10|4      &11|11|5
			\end{tabular}
			\caption{$s=2$, $m=-2$}
		\end{subtable}
	}
	\caption{$|s|=2$, $|m|=2$, purely dynamical initial data}
	\label{tab:s2m2dyn}
\end{table}

\begin{enumerate}
	\item As in other cases, the LPI values at the horizon $R_+$ are always increased at least by $1$ while  the value of $\ell$ is increased by $1$.
	\item The LPI values at future null infinity $\scri^+$ do not necessarily increase, though they never decrease while the value of $\ell$ is increased.
	\item For any fixed value of  $s$, the LPIs are independent of the sign of $m$.
	\item The LPI values $\mu_{R_+}$ at the horizon are independent of the sign of the spin parameter $s$,
	with some exceptions when $\ell'=\ell_0+1$, $\ell\geq\ell'$, $|s|=1$, $m=2$. Unfortunately, we do not have the data to check if a similar effect is present when $|s|=2$. Nevertheless, using also Table 5 of \cite{Harms:2013ib}, it can be seen that the effect is not present for $|s|=2$, $m=2$.
	\item As above, the  LPI values $\mu_{\scri^+}$ do depend on the sign of $s$ such that the value of $\mu_{\scri^+}$ relevant for negative $s$ is always larger by $2\,|s|$ than the LPI value  $\mu_{\scri^+}$ pertinent to the corresponding positive $s$.
\end{enumerate}
Concerning the two Green's-function-based arguments, we draw the following conclusions from our numerical findings:
\begin{enumerate}
  \setcounter{enumi}{5}
	\item The LPI values at the horizon $\mu_{R_+}$ are  reassuring that the modifications in \cite{Harms:2013ib} are valid only in the case $m=0$. A notable exception is the row $\ell'=3$ in Table \ref{tab:m2dyn}-a. There, the data suggest that for $\ell>2$ a similar modification is needed as in the $m=0$ case.
	\item For $s>0$, apart from $\ell'=\ell_0+1$, the values $\mu_{R_+}$ and the LPI values at intermediate locations $R_+<R<1$ are always equal to each other. Whenever $\ell'=\ell_0+1$, our LPI values $\mu_{R}$ at intermediate locations, with the exception of $\ell=\ell_0$,  are systematically larger by $1$ than predicted by (\ref{eq: PD2}).
    Again, this suggests that a correction similar to the one presented in \cite{Harms:2013ib} is needed whenever $s>0$, $\ell'=\ell_0+1$, and $\ell>\ell_0$.
	\item At future null infinity our numerically determined values for $\mu_{\scri^+}$ are consistent with the results of \cite{Hod:2000fh,Harms:2013ib}.
	\end{enumerate}

All in all, it appears that in the very limited subcase with $\ell'=\ell_0+1$, there are certain special circumstances where neither the predictions made in \cite{Hod:2000fh} nor those made in \cite{Harms:2013ib} are supported by our numerical findings. The corresponding subcase would definitely desire more thoughtful analytic inspection.

\subsection{Energy and angular momentum conservation}\label{subsec: energy-balance-relations}

As it was shown in \cite{Toth:2018ybm}, to any pair of spin $s$ and $-s$ solutions of the Teukolsky master equations, there always exist some conserved canonical energy- and angular-momentum-type currents. Since the corresponding vector fields $E^a$ and $J^a$ are divergence free, i.e.,\,$\nabla_a E^a=0$ and $\nabla_aJ^a=0$, and by construction are complex, they provide us two complex balance relations as formulated by \eqref{eq: balance} along with its correspondent yielded by the replacement of $E^a$ by $J^a$.
All in all, the real and imaginary parts of these two complex balance relations give us four real ones, and these four together provide us---in addition to the conventional checks such as the numerical convergence rate---a very important verification of the correctness and robustness of our numerical results.

\medskip

\begin{figure}[ht!]
	\begin{centering}
		{\tiny
			\begin{subfigure}{0.48\textwidth}
				\includegraphics[width=\textwidth]{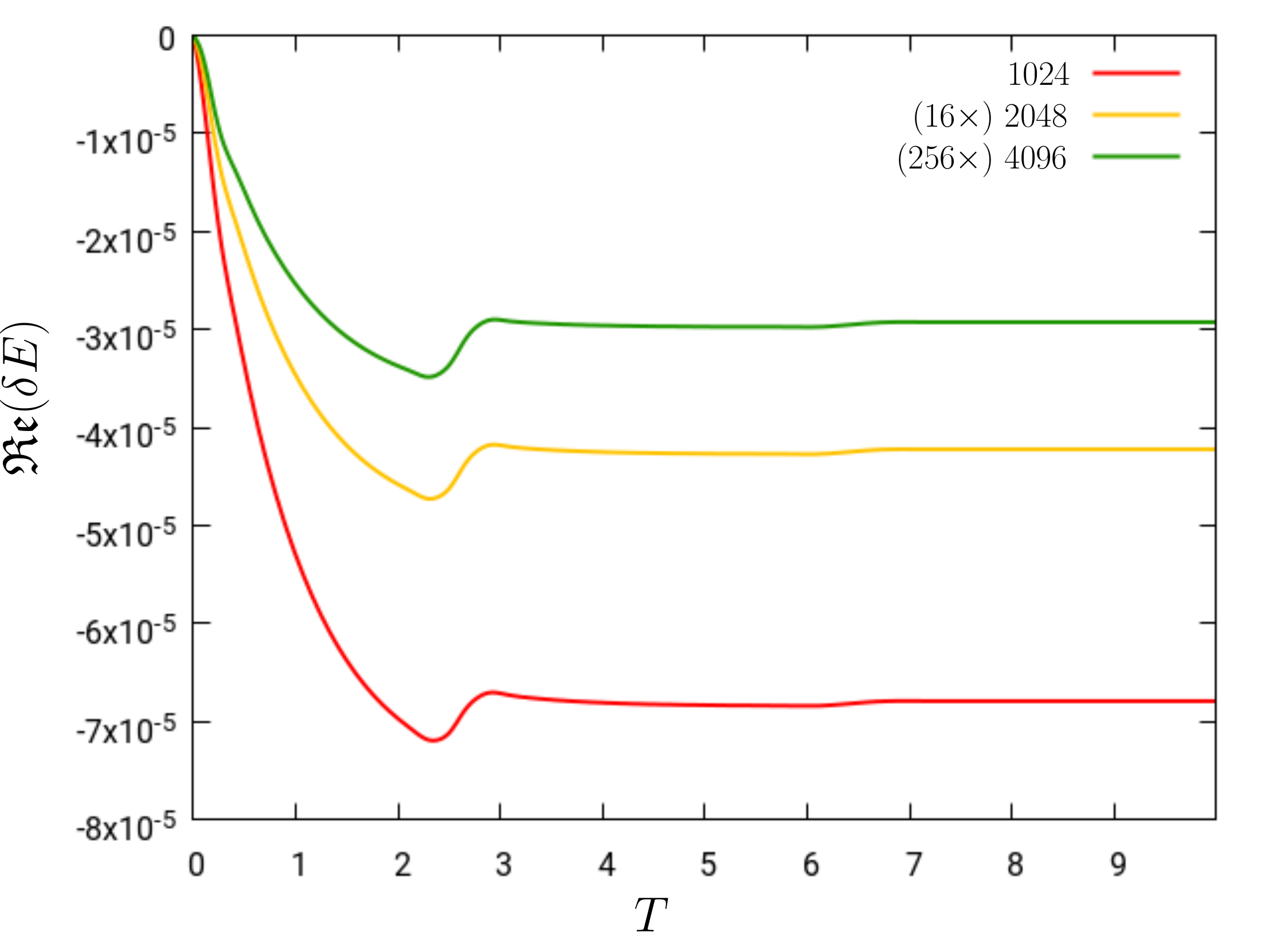}
				\caption{\scriptsize The time dependence of the real part $\mathfrak{Re}(\delta E)$ of energy balance relation $\delta E$ is plotted for three resolutions. The convergence rate is slightly better than $4$.}
				\label{fig:enbalre}
			\end{subfigure}
			\begin{subfigure}{0.48\textwidth}
				\includegraphics[width=\textwidth]{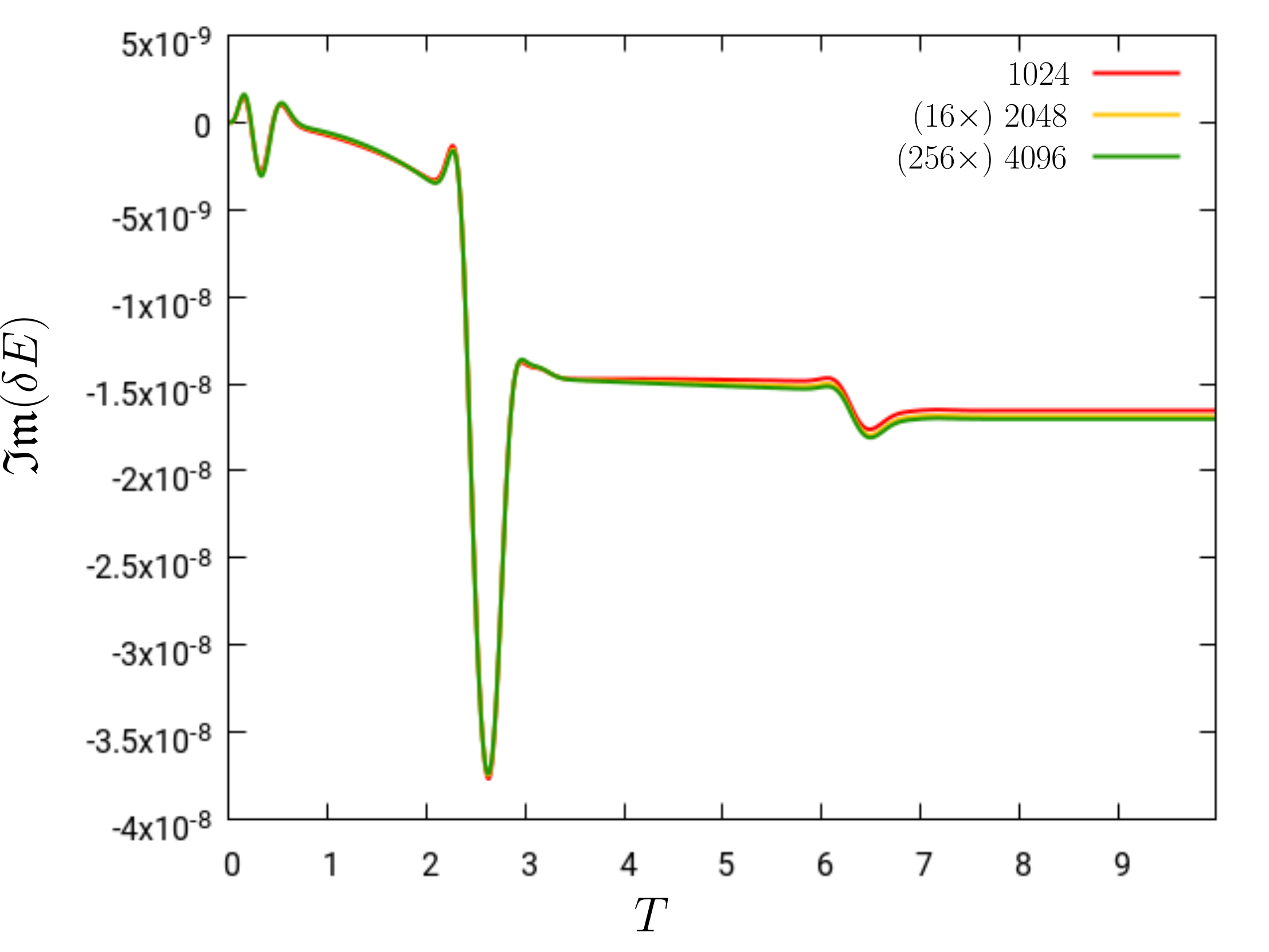}
				\caption{\scriptsize The time dependence of the imaginary part $\mathfrak{Im}(\delta E)$ of energy balance relation $\delta E$ is depicted for three resolutions. The convergence rate is exactly $4$. }
				\label{fig:enbalim}
			\end{subfigure}
			\begin{subfigure}{0.48\textwidth}
				\includegraphics[width=\textwidth]{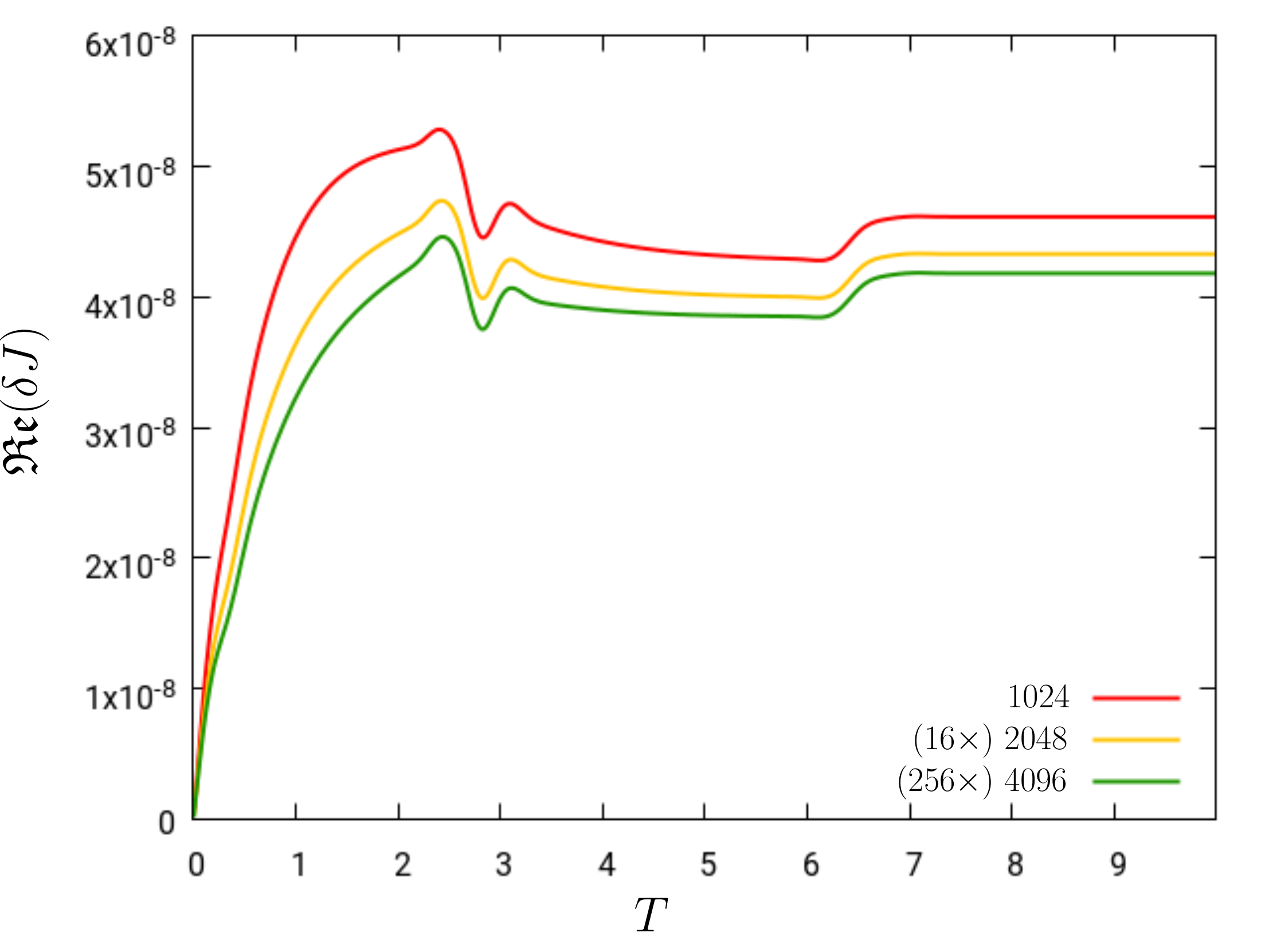}
				\caption{\scriptsize The time dependence of the real part $\mathfrak{Re}(\delta J)$ of angular momentum balance relation $\delta J$ is shown. The convergence rate is seen to be slightly better than $4$.}
				\label{fig:angbalre}
			\end{subfigure}\hskip.5cm
			\begin{subfigure}{0.48\textwidth}
				\includegraphics[width=\textwidth]{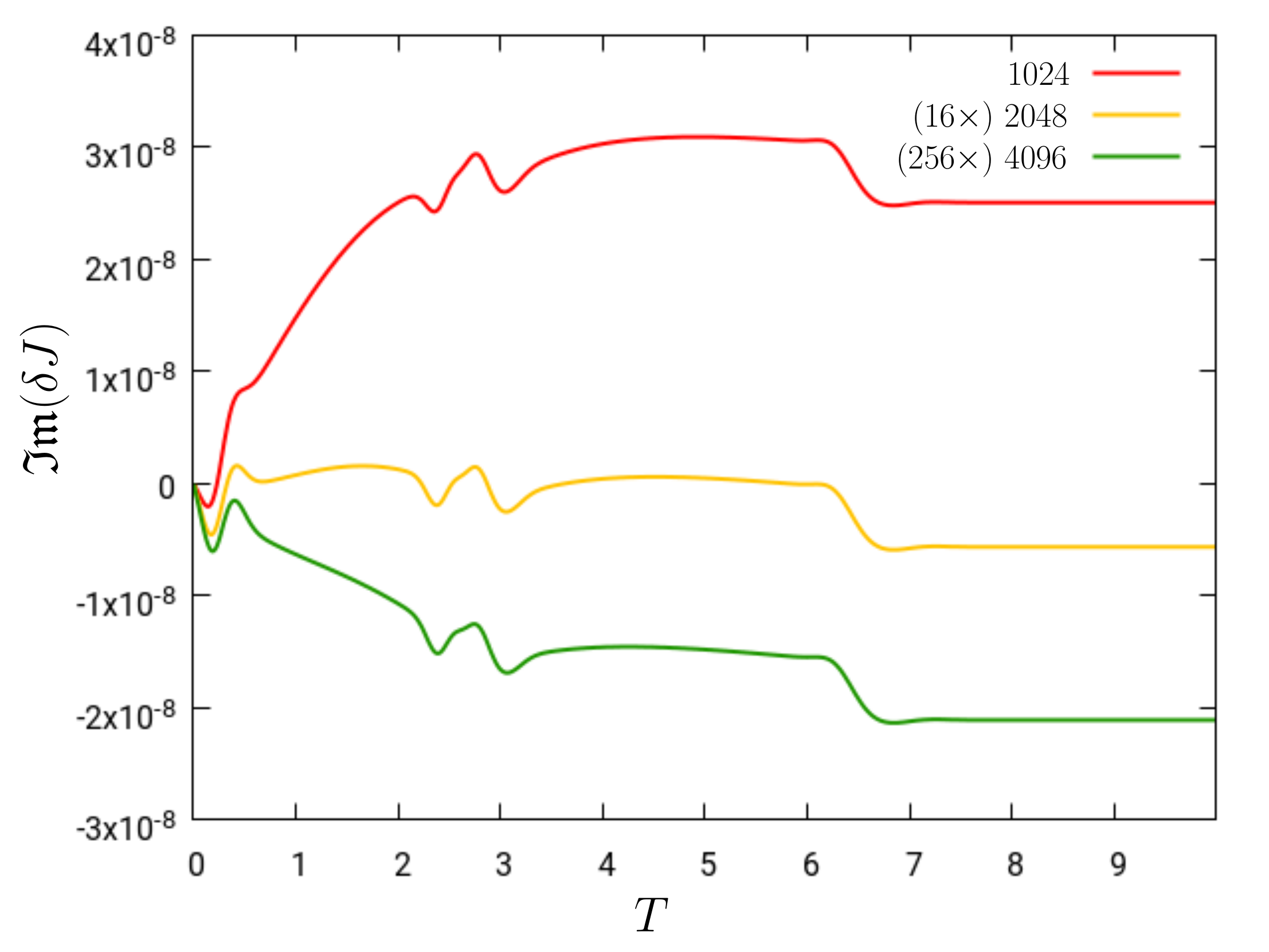}
				\caption{\scriptsize The time dependence of the imaginary part $\mathfrak{Im}(\delta J)$ of angular momentum balance relation $\delta J$ is shown. The convergence rate remains close to $4$.}
				\label{fig:angbalim}
			\end{subfigure}
		}
	\end{centering}
	\caption{\small The time dependence of the real and imaginary parts of the energy and angular momentum balance relations $\delta E$ and $\delta J$ are shown, respectively. This is done in each panel for three different resolutions with $1024, 2048, 4096$ spatial grid points.  In order to make it transparent that our numerical implementation, as desired, is of fourth order accurate, the numerical values of $\delta E$ and $\delta J$ relevant for the resolutions $2048$ and $4096$ are multiplied by $16$ and $256$, respectively.}
	\label{fig:convbal}
\end{figure}

In this subsection, first these energy and angular momentum balance relations will be applied. In particular, in the panels of Fig.\,\ref{fig:convbal}, the time dependence of the real and imaginary parts of the numerical errors
\begin{equation}\label{eq: balance-numerical}
\delta E=\int_{\partial\Omega}n_\mu E^\mu \qquad \mathrm{and} \qquad \delta J=\int_{\partial\Omega}n_\mu J^\mu
\end{equation}
are plotted against time. The applied rectangular domain of integration $\Omega$ is bounded by an initial  time slice $T=T_i$ and by a running $T$ slice,
and also by the null cylinders at the horizon $R=R_{+}$ and at future null infinity at $R=1$, respectively. Note that on analytic grounds, the integrals $\delta E$ and $\delta J$ in the  balance relations in \eqref{eq: balance-numerical} should vanish identically. Accordingly, they have to be (and they are expected to stay) small if the numerical implementation is correct.
It is also important to emphasize that in each panel of Fig.\,\ref{fig:convbal}, the aforementioned time dependences are plotted for three different resolutions, with $1024, 2048, 4096$ spatial grid points. In this respect, the panels in Fig.\,\ref{fig:convbal}  are not only to demonstrate that the real and imaginary parts of the balance relations \eqref{eq: balance-E} and \eqref{eq: balance-J} hold at a satisfactorily accurate level but also to make it transparent that our numerical implementation has a fourth order convergence rate---even in the considered highly complicated expressions deduced from the numerically determined basic variables---as it should happen as a fourth order accurate Runge-Kutta time integrator is applied in the $T-R$ section. To make this transparent, the numerical values of $\delta E$ and $\delta J$ relevant for the resolutions $2048$ and $4096$ are multiplied by $16$ and $256$, respectively.

If the numerical values of $\delta E$ and $\delta J$ were exactly proportional to the fourth power of the grid spacing $\delta R$, then the lines corresponding to the three different resolutions shown in the panels of Fig.\,\ref{fig:convbal} would exactly coincide. The apparent deviation from this prediction, which is very small in Fig.\,\ref{fig:enbalim} but larger in the other three panels,
can be explained by higher order corrections to the leading $\propto(\delta R)^4$ power law.
In particular, the assumption that the dependence of the numerical values of $\delta E$ and $\delta J$ on $\delta R$
has the functional form $c_4(\delta R)^4+c_5(\delta R)^5+\dots$,
where $c_4$ and $c_5$ are suitable constants, implies that
the distance between the lines for the $2048$ and $4096$ grid points should be approximately half the distance
between the lines for the $1024$ and $2048$ grid points, which can indeed be seen in Fig.\,\ref{fig:convbal}.

\medskip

The conserved currents $E^a$ and $J^a$ can also be used to analyze transport in the energy- and angular-momentum-type quantities, and we shall do so in the rest of this subsection. Nevertheless, in advance of doing so it is important to emphasize that these currents, apart from the spin-weight $s=0$ case, are not defined for individual fields but only for a pair of spin-weight $s$ and $-s$ solutions to the Teukolsky master equation. This restriction applies even though the spin-weight $s$ and $-s$ solutions can be generic; i.e.,~they could completely be independent of each other.

\begin{figure}[ht!]
 \begin{centering}
 {\tiny
 \begin{subfigure}{0.48\textwidth}
 \includegraphics[width=\textwidth]{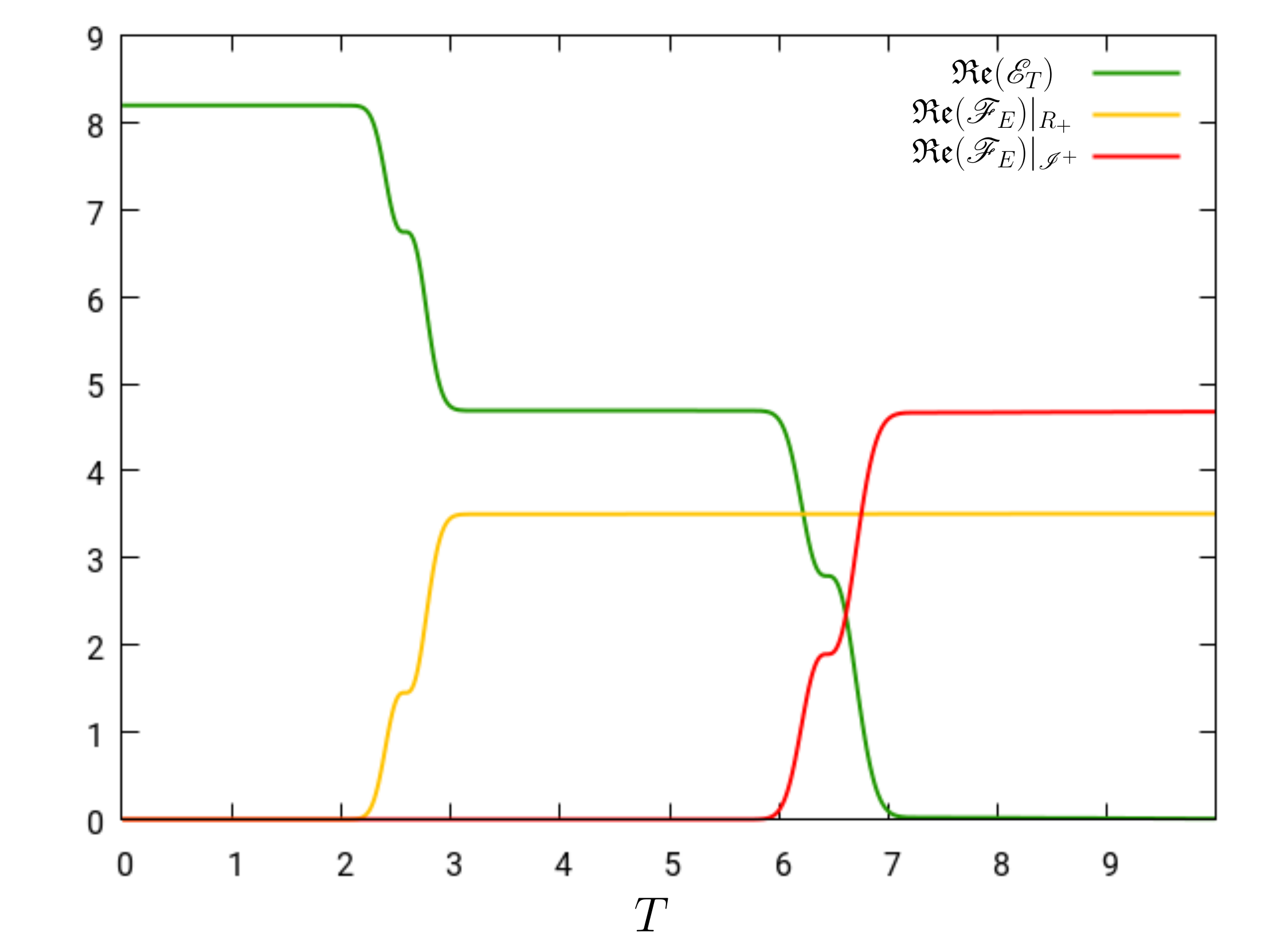}
 \caption{\scriptsize The time dependence of the real part $\mathfrak{Re}(\mathscr{E}_T)$ of the total energy $\mathscr{E}_T$, along with the real part $\mathfrak{Re}(\mathscr{F}_E)$ of the integrated energy flow $\mathscr{F}_E$, evaluated at $R_+$ and at $\scrip$, is plotted. It is visible that first about half of the energy stored by the initial data is leaving the domain of outer communication via the black hole event horizon. The other half leaves later through future null infinity $\scrip$.}
 \label{fig:enbegre}
 \end{subfigure}
 \begin{subfigure}{0.48\textwidth}
 \includegraphics[width=\textwidth]{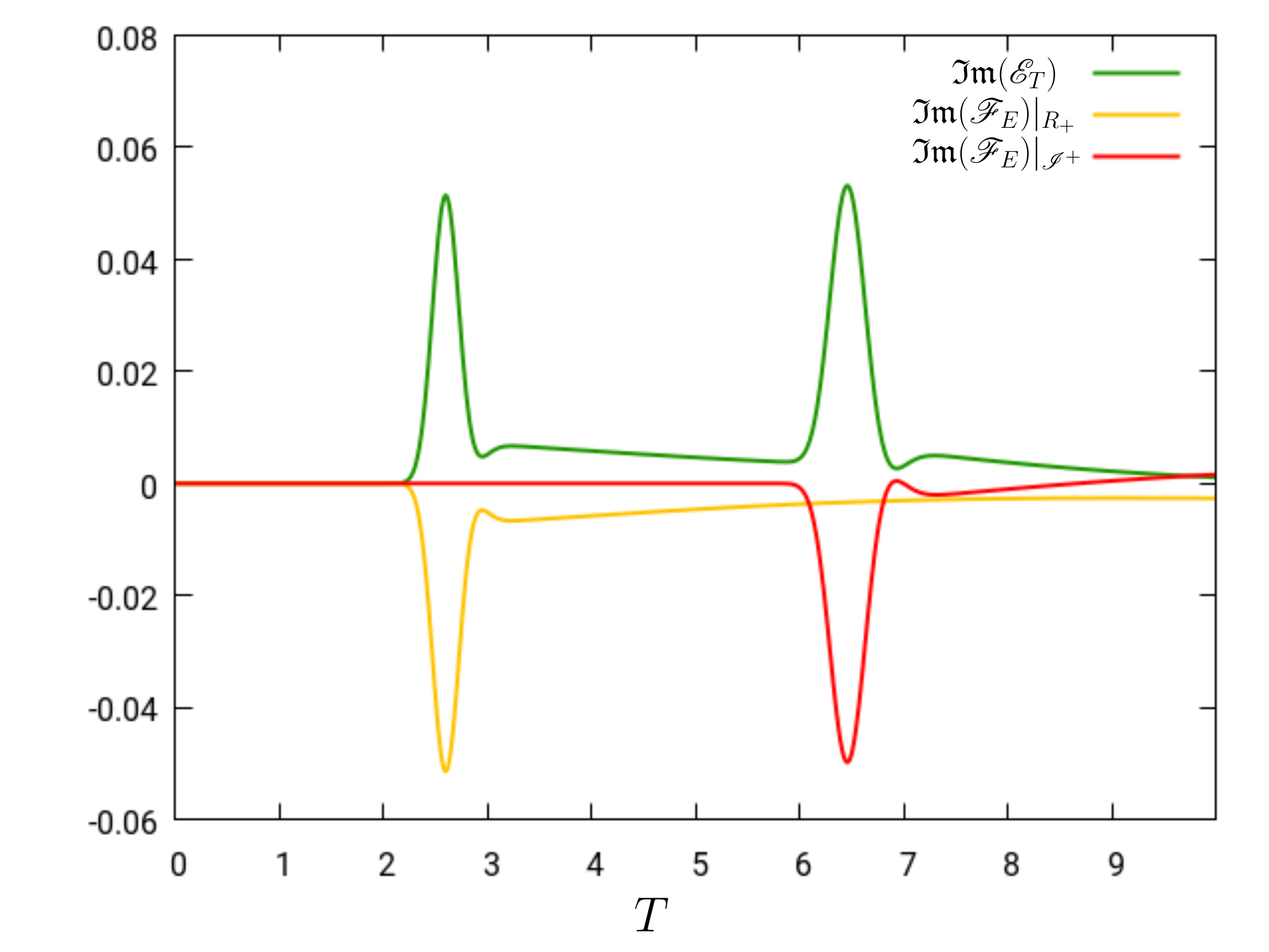}
 \caption{\scriptsize The imaginary part $\mathfrak{Im}(\mathscr{E}_T)$ of the total energy $\mathscr{E}_T$, along with the imaginary part $\mathfrak{Im}(\mathscr{F}_E)$ of the integrated energy flow $\mathscr{F}_E$, evaluated at $R_+$ and at $\scrip$, is plotted against time. Two spikes are  visible on this figure which occur exactly when the pulses go through the event horizon and through future null infinity. Apart from these spikes the transport processes here remain small scale.}
 \label{fig:enbegim}
 \end{subfigure}
 \begin{subfigure}{0.48\textwidth}
 \includegraphics[width=\textwidth]{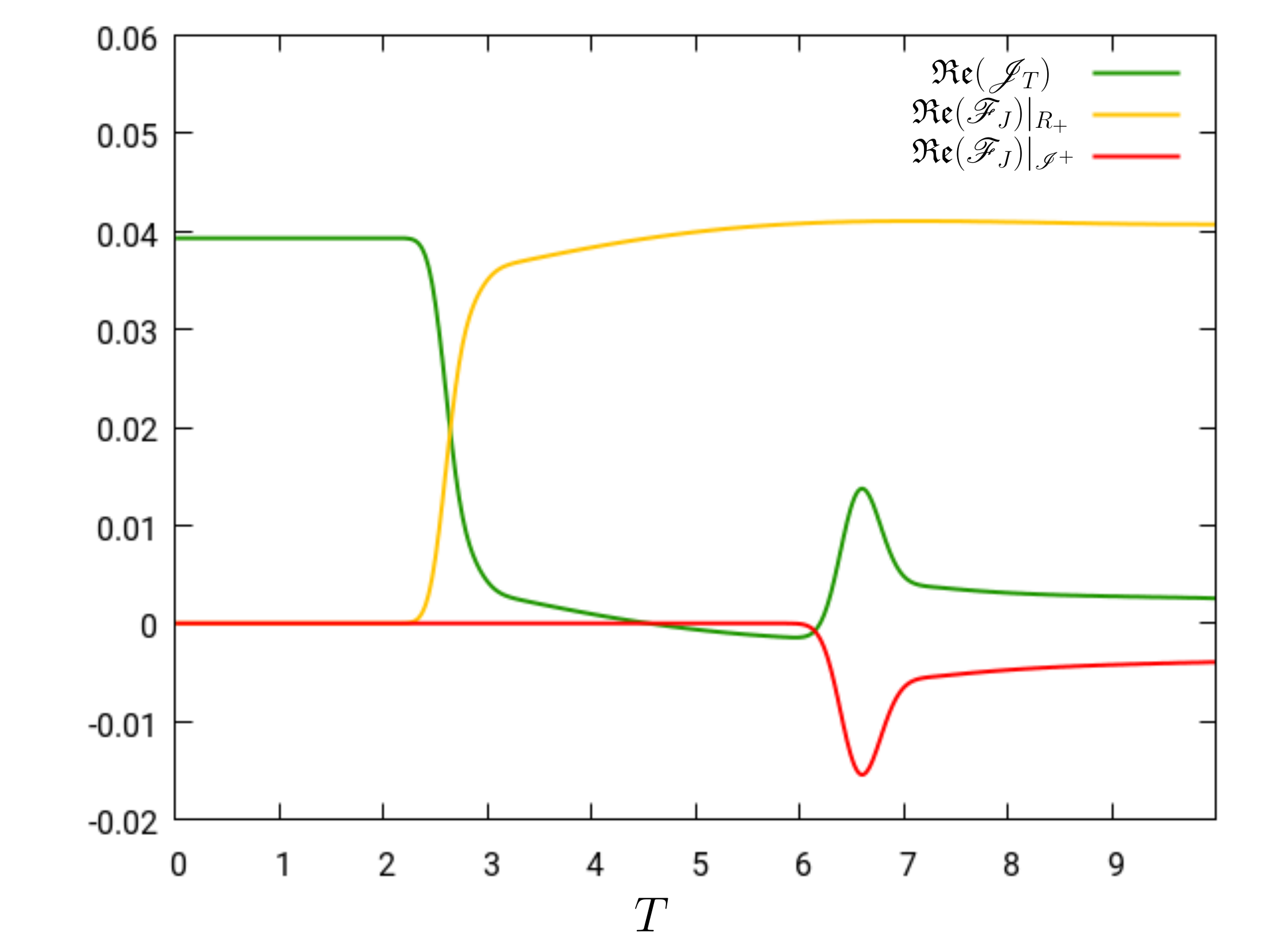}
 \caption{\scriptsize The time dependence of the real part $\mathfrak{Re}(\mathscr{J}_T)$ of the total angular momentum $\mathscr{J}_T$, along with the real part $\mathfrak{Re}(\mathscr{F}_J)$ of the integrated energy flow $\mathscr{F}_J$, evaluated at $R_+$ and at $\scrip$, is plotted. The angular momentum drops significantly by the loss through the black hole event horizon which is followed by some negative flow through null infinity. Notably, there is more angular momentum in the system at $T=10$ than around $T=5$.}
 \label{fig:angbegre}
 \end{subfigure}\hskip.5cm
 \begin{subfigure}{0.48\textwidth}
 \includegraphics[width=\textwidth]{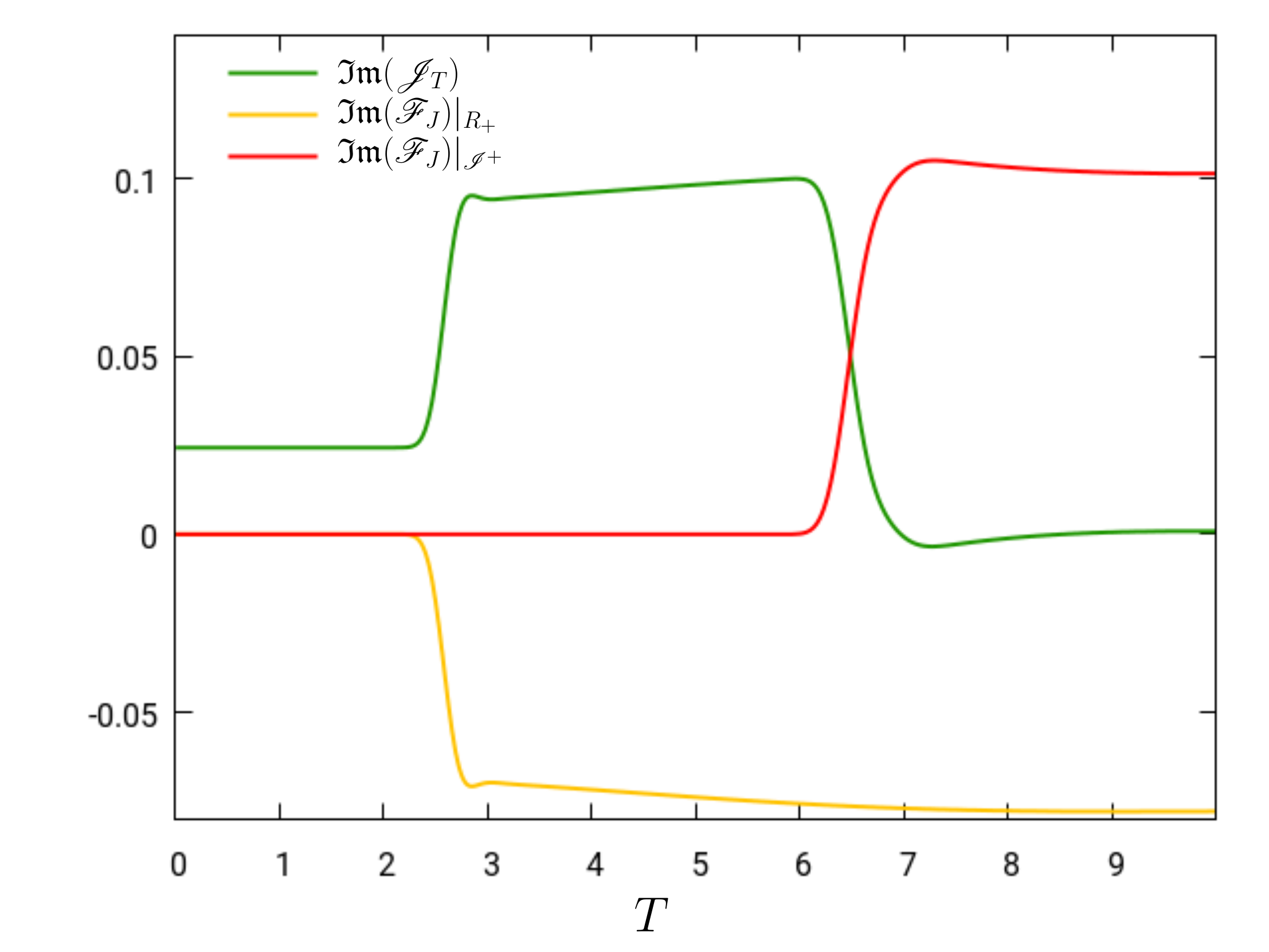}
 \caption{\scriptsize The imaginary part $\mathfrak{Im}(\mathscr{J}_T)$ of the total angular momentum $\mathscr{J}_T$, along with the imaginary part $\mathfrak{Im}(\mathscr{F}_J)$ of the integrated energy flow $\mathscr{F}_J$, evaluated at $R_+$ and at $\scrip$, is plotted against time. The losses here are of opposite sign w.r.t. the real part; i.e.,~the flow through the black hole event horizon is negative whereas the flow through null infinity is positive. These losses almost emptying the imaginary part of the total angular momentum.}
 \label{fig:angbegim}
 \end{subfigure}
 }
 \end{centering}
 \caption{\small The time dependence of the real and imaginary parts of the total energy $\mathscr{E}_T$ and total angular momentum $\mathscr{J}_T$, as well as those of the corresponding flows $\mathscr{F}_E$ and $\mathscr{F}_J$ are depicted. Though only the initial parts of the time evolution are shown in these plots, they make transparent about $99\%$ of the pertinent transport processes. }
 \label{fig:dens}
 \end{figure}

In the panels in Fig.\,\ref{fig:dens},
the time dependence of the real and imaginary parts of the total energy and angular momentum,
\begin{equation}
\mathscr{E}_T=\int_{\Sigma_T}n_a^{(T)}E^a\sqrt{|h_T|}\,\,\dd R\,\dd\vartheta\,\dd\varphi \quad \mathrm{and} \quad \mathscr{J}_T=\int_{\Sigma_T}n_a^{(T)}J^a\sqrt{|h_T|}\,\,\dd R\,\dd\vartheta\,\dd\varphi\,,
\end{equation}
where $\Sigma_T$ signifies the $T=const$ slices, as well as the time dependence of the integrals of the corresponding flows
\begin{equation}
\mathscr{F}_E=\int_{{{Cyl}}_{R,T}} \hskip-0.5cm n_a^{(R)}E^a\sqrt{|h_R|}\,\,\dd T\,\dd\vartheta\,\dd\varphi \quad \mathrm{and} \quad \mathscr{F}_J=\int_{{{Cyl}}_{R,T}} \hskip-0.5cm n_a^{(R)}J^a\sqrt{|h_R|}\,\,\dd T\,\dd\vartheta\,\dd\varphi\,,
\end{equation}
where the latter integrals are supposed to be evaluated at the cylinders $R_+\times [T_i,T]$ at the black hole event horizon and $R_{\scrip}\times [T_i,T]$  at future null infinity, are plotted, respectively. These plots are showing only the initial parts of the time evolution; nevertheless, as $99\%$ of the transport processes happens in the corresponding initial intervals, these plots are informative about the characteristic features.

\medskip

Even though the spin-weight $s$ and $-s$ solutions to the Teukolsky master equation may be completely independent in our simulations, they were yielded by applying exactly the same $R$-dependent bumpy profile \eqref{eq: initial-data-bump}, with exactly the same parameters as described in Sec. \ref{subsec: initial-data}. Probably for this reason, there is a striking similarity between the real parts of the energy- and angular-momentum-type transport processes, as depicted in Figs. \ref{fig:dens}(a) and \ref{fig:dens}(c) and between the corresponding figures (5) and (7) in \cite{Racz:2011qu} depicting the true energy and angular momentum transport processes relevant for a single spin-weight zero scalar field investigated in detail in \cite{Racz:2011qu}.

\medskip

In closing this subsection, note that it is obviously interesting to know what happens with the conserved currents and with their integrals under the flip $a \mapsto -a$ of the sign of the rotation parameter of the background Kerr spacetime. Remarkably, only the following two flippings are induced by this transformation.
Notably, a flipping happens in the signs of the imaginary part $\mathfrak{Im}(\mathscr{E}_T)$ of the total energy $\mathscr{E}_T$ and with the imaginary part $\mathfrak{Im}(\mathscr{F}_E)$ of the integrated energy flow $\mathscr{F}_E$ evaluated at $R_+$ and at $\scrip$, whereas a completely analogous flipping happens in the real part $\mathfrak{Re}(\mathscr{J}_T)$ of the total angular momentum  $\mathscr{J}_T$ and with the real part  $\mathfrak{Re}(\mathscr{F}_J)$ of the integrated energy flow $\mathscr{F}_J$ evaluated at $R_+$ and at $\scrip$.
The sign flips in the real part of the angular momentum integrals seem to be in accordance with intuition as under the flip $a \mapsto -a$ of the rotation parameter the background Kerr black hole gets to be counterrotating with respect to the spin $s$ and $-s$  solutions of the Teukolsky master equation.

\section{Final remarks}\label{sec: final-results}

The time evolution and the late-time behavior of spin $s$ fields with $s=\pm 1, \pm 2$ on a fixed Kerr background was examined numerically. The applied mathematical setup incorporated the techniques of conformal
compactification and the hyperbolic initial value formulation of these spin $s$ fields. A new code was introduced that was developed based on the two-parameter  foliation of the Kerr background by topological two-spheres as determined by the Boyer-Lindquist $t=const$ and $r=const$ level surfaces. The angular dependences along the foliating two-surfaces were treated by applying multipole expansions of the basic variables in terms of spin-weighted spherical harmonics, by which the applied method gets to be fully spectral in these angular directions. In the complementary time-radial section, the method of lines in a fourth order accurate  Runge-Kutta time integration scheme was applied.

\medskip

The time evolution of purely static and purely dynamical initial data were investigated.
The asymptotic decay rates were evaluated at the black hole event horizon, in the domain
of outer communication and at future null infinity. This was done systematically by scanning through a significantly wide range of the input and output parameters $\ell', m, s, \ell$. The deduced decay rates were compared to those which were deduced prior to our investigation. In particular,  our most important findings can be summarized as follows by referring to the results covered by \cite{Harms:2013ib,Hod:1999ci,Hod:1999rx,Hod:2000fh}.

\medskip

As noted earlier, in the case of static initial data, we compared our numerical results with those deduced also numerically  in \cite{Harms:2013ib}. The only limitation in doing so was that (apart from a few exceptions) the decay exponents were determined in \cite{Harms:2013ib} only for axially symmetric configurations with $m=0$.
Remarkably, for purely static initial data, if $|s|$ is replaced by $\ell_0=\max\{|m|,|s|\}$ everywhere in \eqref{eq: ST1} and \eqref{eq: ST2}---these were deduced in \cite{Harms:2013ib} for the $m=0$ case---the yielded relations [see \eqref{eq: ST1-own} and \eqref{eq: ST2-own}  below] are automatically valid to the fully general (not necessarily axially symmetric) configurations.
Accordingly, for static configurations, our findings---these are collected in Tables \ref{tab:axisymm-s1}-a, \ref{tab:axisymm-s1}-b, \ref{tab:axisymm-s2}-a, \ref{tab:axisymm-s2}-b, for axisymmetric configurations with $m=0$ and  in Tables  \ref{tab:m1stat}--\ref{tab:s2m2stat} for nonaxisymmetric configurations with $m\not=0$---can be summarized as follows:
\begin{enumerate}
	\item[(ST1)] At the horizon $R=R_+$
	\begin{equation}\label{eq: ST1-own}
	\hskip0.8cm
	n|_{R_+} = \begin{cases}
	\ell'+\ell+3+\alpha\,,& \mbox{if}\ \ \ell'=\ell_0\,,\\
	\ell'+\ell+2+\alpha\,,&\mbox{if }\ \ \ell'>\ell_0\,,
	\end{cases}
	\end{equation}
	where---now as it was in \eqref{eq: PD1}---$\ell_0=\max\{|m|,|s|\}$ and $\alpha=0$ in all cases except if $s>0$ and $m=0$, when $\alpha=1$.

	\item[(ST2)] At any finite intermediate spatial location with $R_+<R<1$,
	\begin{equation}\label{eq: ST2-own}
	n|_{R} = \begin{cases}
	\ell'+\ell+3\,,& \mbox{if}\ \ \ell'=\ell_0\,,\\
	\ell'+\ell+2\,,&\mbox{if }\ \ \ell'>\ell_0\,.
	\end{cases}
	\end{equation}

	\item[(ST3)] Finally, at future null infinity signified by $R=1$,
	\begin{equation}\label{eq: ST3-own}
	\hskip-0.3cm
	n|_{R=1} = \begin{cases}
	\ell-s+2\,,& \mbox{if}\ \ \ell'\leq \ell\,,\\
	\ell'-s+1\,,&\mbox{if }\ \ \ell' > \ell\,.
	\end{cases}
	\end{equation}
\end{enumerate}

As for the solutions of the Teukolsky master equation starting with purely dynamical initial data, as indicated earlier, the situation gets more involved at least in certain subcases.
The relevant data are collected in Tables \ref{tab:axisymm-s1}-c, \ref{tab:axisymm-s1}-d, \ref{tab:axisymm-s2}-c, \ref{tab:axisymm-s2}-d for axisymmetric and in Tables \ref{tab:m1dyn}-\ref{tab:s2m2dyn} for nonaxisymmetric initial data.

\begin{enumerate}
	\item[(PD1)] For instance, at the horizon with $R=R_+$, for $m\neq 0$ and $\ell'=\ell_0+1$ the decay rates based on our numerical findings do not fully agree with the former predictions. Our results can be described by the following
	modification of (\ref{eq: PD1}):
    \begin{equation}
        n|_{R_+} = \begin{cases}
    	\ell'+\ell+3+\alpha\,,& \mathrm{if}\ \ \ell'=\ell_0\,,\\
        \ell'+\ell+3+\alpha-\widetilde{\delta}\,,& \mathrm{if}\ \ \ell'=\ell_0+1\,,\\
    	\ell'+\ell+1+\alpha\,,&\mathrm{if}\ \ \ell'>\ell_0+1\,,
    	\end{cases}
	\end{equation}
    where $\ell_0=\max\{|m|,|s|\}$, $\alpha=0$ in all cases except if $s>0$ and $m=0$, when $\alpha=1$, and also $\tilde{\delta}=0$ if $m=0$ or $m=2$ with $s=-1$ and $\ell>\ell_0$, $\tilde{\delta}=1$ in all other cases.

	\item[(PD2)] At finite intermediate spatial locations with $R_+<R<1$, we also found that there are differences between
	(\ref{eq: PD2}) and our results if $\ell'=\ell_0+1$ and $m\neq 0$.
	Again, this difference can be taken into account by a
	minor modification of $\delta$:
    \begin{equation}
        n|_{R} = \begin{cases}
    	\ell'+\ell+3+\alpha\,,& \mathrm{if}\ \ \ell'=\ell_0\,,\\
        \ell'+\ell+3+\alpha-\widehat{\delta}\,,& \mathrm{if}\ \ \ell'=\ell_0+1\,,\\
    	\ell'+\ell+1+\alpha\,,&\mathrm{if}\ \ \ell'>\ell_0+1\,,
    	\end{cases}
	\end{equation}
    where $\ell_0$ and $\alpha$ are as above, and also $\widehat{\delta}=0$ if $m=0$ or $m\neq 0$ but $\ell>\ell_0$ and $\widehat{\delta}=1$ otherwise.

	\item[(PD3)] At $R=1$ representing future null infinity $\mathscr{I}^+$,
	all of our pertinent numerical findings support the predictions of (\ref{eq: PD3}); i.e.,~the following rules apply
	\begin{equation}\label{eq: PD3-own}
	\hskip-2.55cm
	n|_{R=1} = \begin{cases}
	\ell-s+2+\gamma\,,& \mathrm{if}\ \ \ell'\leq \ell+1\,,\\
	\ell'-s\,,&\mathrm{if}\ \ \ell' > \ell+1\,,
	\end{cases}
	\end{equation}
	where  $\gamma=0$ in all cases except if $m=0$, $\ell'=\ell_0+1$, and $\ell=\ell_0$, when $\gamma=1$.
\end{enumerate}

The apparent dependence of the decay rate on the value of the azimuthal parameter $m$ at the horizon $R=R_+$ and intermediate locations $R_+<R<1$ indicate that there may be interesting features which will definitely desire further investigation. To clear this up, along with possibly some other interesting issues is, however, out of the scope of the present paper and they are left for future studies.



\section*{Acknowledgments}


K. C. and G. Z. T., were supported in part by the NKFIH Grants No. K-115434 and No. K-116505, respectively.
I. R. was supported by the POLONEZ program of the National Science Centre of Poland (under Project No. 2016/23/P/ST1/04195) which has received funding from the European Union's Horizon 2020 research and innovation program under the Marie Sk{\l}odowska-Curie Grant Agreement No.~665778.

\includegraphics[scale=0.42, height = 42pt]{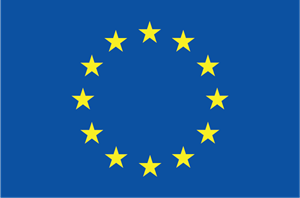}

\newpage

\section*{Appendixes}
\appendix

The succeeding appendixes are to provide the explicit form of the expressions mentioned and applied in the main part of the discussions of the foregoing sections. Also, we give a brief summary of the use of spin-weighted spherical harmonics.

\section{The homogeneous TME}
 \label{app:TME_coeff}
\renewcommand{\theequation}{A.\arabic{equation}}
\setcounter{equation}{0}

As mentioned in Sec. \ref{subsec: regularization}, after regularization, the homogeneous Teukolsky master equation takes the form
\begin{align}
&\diff_{TT}\Phi^{(s)}=\frac{1}{\mathscr{A}+\mathscr{B}\cdot Y_2^0}\,\Big[c_{RR}\cdot\diff_{RR}\Phi^{(s)}+c_{TR}\cdot\diff_{TR}\Phi^{(s)}+c_{T\varphi}\cdot\diff_{T\varphi}\Phi^{(s)}+c_{R\varphi}\cdot\diff_{R\varphi}\Phi^{(s)} \nonumber\\
&+c_{\vartheta\vartheta}\cdot\overline{\eth}\eth\,\Phi^{(s)}+c_T\cdot\diff_T\Phi^{(s)}+\mathrm{i}\,c_{Ty} \,Y_1^0\cdot\diff_T\Phi^{(s)}+c_R\cdot\diff_R\Phi^{(s)}+c_\varphi\cdot\diff_\varphi\Phi^{(s)}+c_0\cdot\Phi^{(s)}\Big]\,.\nonumber
\end{align}
All the involved $R$-dependent coefficients are listed below. In particular, $\mathscr{A}$ and $\mathscr{B}$ are
 \begin{equation}
 \mathscr{A}=\mathfrak{a}+\mathfrak{b}/3\qquad\text{and}\qquad \mathscr{B}=\frac{4}{3}\sqrt{\frac{\pi}{5}}\,\mathfrak{b}\,,\qquad\text{where}
 \end{equation}
 \vskip-0.4cm
 \begin{align}
 \mathfrak{a} =& -4R \,\Big(\,A_0 + A_1R + A_2R^2 + A_3R^3 + A_4R^4 + A_5R^5 + A_6R^6\Big)\,,\\
  &A_0 = -M\,,\\
  &A_1 = a^2 - 1 + 4a^2M - 5M + 4a^2M^2 - 8M^2\,,\\
  &A_2 = a^2 - 1 + 4a^2M - 11M + 4a^2M^2 - 24M^2 - 16M^3\,,\\
  &A_3 = -4a^2M + M - 8a^2M^2 - 8M^2 - 16M^3\,,\\
  &A_4 = -4a^2M - 8a^2M^2 + 8M^2 + 16M^3\,,\\
  &A_5 = 4a^2M^2 + 16M^3\,,\\
  &A_6 = 4a^2M^2\,,\\[2mm]
  \mathfrak{b} =& \ a^2(1 + R)(R^2+1)^2.
  \end{align}
The rest of the involved coefficients are
 \begin{align}
  c_{RR} =& \ \frac{1}{4}(1 + R)(R^2-1)^2(c_{RR0} + c_{RR1}R + c_{RR2}R^2 + c_{RR3}R^3 + c_{RR4}R^4)\,,\\
  &c_{RR0} = a^2\,,\\
  &c_{RR1} = -4M\,,\\
  &c_{RR2} = -2(a^2 - 2)\,,\\
  &c_{RR3} = 4M\,,\\
  &c_{RR4} = a^2\,,\\[6mm]
  c_{TR} =& \  2\,R\,(1 + R)\,\Big[\,c_{TR0} + c_{TR1}R + c_{TR2}R^2 + c_{TR3}R^3 + c_{TR4}R^4 + c_{TR5}R^5 + c_{TR6}R^6\,\Big]\,,\\
  &c_{TR0} = -a^2 - 2a^2M + 2M\,,\\
  &c_{TR1} = 4M(1 + 2M)\,,\\
  &c_{TR2} = 2(a^2 - 2 + 3a^2M - 4M)\,,\\
  &c_{TR3} = -4M(1 + 4M)\,,\\
  &c_{TR4} = -a^2 - 6a^2M + 6M\,,\\
  &c_{TR5} = 8M^2\,,\\
  &c_{TR6} = 2a^2M\,,  \\[3mm]
  c_{T\varphi} =&\  4aR(1 + R)(1 + R^2)(-1 - 2M + 2MR^2)\,,\\
  c_{R\varphi} =& \ a(1 + R^2)(R-1)^2(R+1)^3\,,\\
 c_{\vartheta\vartheta} =& \ (1 + R)(R^2+1)^2\,,
 \end{align}
 \begin{align}
 c_T =& \ \frac{1}{1 + R^2}\Big(c_{T0} + c_{T1}R + c_{T2}R^2 + c_{T3}R^3 \nonumber \\
 &\hskip2.2cm + c_{T4}R^4 + c_{T5}R^5 + c_{T6}R^6 + c_{T7}R^7 + c_{T8}R^8 + c_{T9}R^9\Big)\,,\\
&c_{T0} = a^2 + 2a^2M - 2M - 2Ms\,,\\
&c_{T1} = a^2 + 2a^2M - 2M - 4s - 6Ms - 8M^2s\,,\\
&c_{T2} = 5a^2 - 4 + 12a^2M - 14M + 4s + 6Ms - 8M^2s\,,\\
&c_{T3} = 5a^2 - 4 + 12a^2M - 30M - 48M^2 - 8s + 2Ms - 8M^2s\,,\\
&c_{T4} = -5a^2 + 4 - 24a^2M + 10M - 48M^2 + 8s + 18Ms - 8M^2s\,,\\
&c_{T5} = -5a^2 + 4 - 24a^2M + 26M + 32M^2 - 4s + 22Ms + 8M^2s\,,\\
&c_{T6} = -a^2 + 4a^2M + 6M + 32M^2 + 4s + 10Ms + 8M^2s\,,\\
&c_{T7} = -a^2 + 4a^2M + 6M + 16M^2 + 14Ms + 8M^2s\,,\\
&c_{T8} = 2M(3a^2 + 8M + 4Ms)\,,\\
& c_{T9} = 6a^2M\,,\\
 c_{Ty} =& -4a(1 + R)(R^2+1)^2s\sqrt{\frac{\pi}{3}}\,,
\end{align}

 \begin{align}
  c_R =& \ \frac{1}{2R(1 + R^2)}\big(R-1\big)\big(R+1\big)^2\Big(c_{R0} + c_{R1}R + c_{R2}R^2 + c_{R3}R^3+ c_{R4}R^4 \nonumber \\
  & \hskip 5.8cm+ c_{R5}R^5 + c_{R6}R^6 + c_{R7}R^7 + c_{R8}R^8\Big)\,,\\
 &c_{R0} = a^2\,,\\
 &c_{R1} = -2M(1 + s)\,,\\
 &c_{R2} = 3a^2 + 4s\,,\\
 &c_{R3} = -2M(7 + s)\,,\\
 &c_{R4} = -7a^2 + 12 + 8s\,,\\
 &c_{R5} = 2M(5 + s)\,,\\
 &c_{R6} = a^2 + 4 + 4s\,,\\
 &c_{R7} = 2M(3 + s)\,,\\
  &c_{R8} = 2a^2\,,\\
  c_\varphi =& \ \frac{1}{R}a(R-1)(R^2+1)^2(R+1)^2\,,\\
   c_0 =& \ \frac{1}{2R^2}(R-1)(R^2+1)^2(R+1)^2(-a^2 + 2MR + a^2R^2 + 2MRs)\,.
 \end{align}

\section{Spin-weighted spherical harmonics}\label{app: eth-bareth}
\renewcommand{\theequation}{B.\arabic{equation}}
\setcounter{equation}{0}

One of the advantages of applying the multipole expansion
\begin{equation}
\Phi^{(s)}(T,R,\vartheta,\varphi) = \sum_{\ell=|s|}^{\infty}\sum_{m=-l}^{l}\phi_\ell{}^{m}(T,R)\cdot {}_s{Y_\ell}{}^m(\vartheta,\varphi)
\end{equation}
in terms of spin-weight $s$ spherical harmonics ${}_s{Y_\ell}{}^m$ is that all the angular derivatives
in \eqref{eq:teurt} can be evaluated analytically.
This is done by making use of the $\eth$ and $\overline{\eth}$ operators which act on a function $f$ of spin-weight $s$ as
\begin{align}
\eth f&=-\sin^s\!\vartheta\left(\partial_\vartheta+\frac{\mathrm{i}}{\sin\!\vartheta}\partial_\varphi\right)(\sin^{-s}\!\vartheta \cdot f)\,,\\
\overline{\eth} f&=-\sin^{-s}\!\vartheta\left(\partial_\vartheta-\frac{\mathrm{i}}{\sin\!\vartheta}\partial_\varphi\right)(\sin^{s}\!\vartheta\cdot f)\,.
\end{align}
It is straightforward to check that for their commutator, the relation
\begin{equation}
(\overline{\eth}\eth-\eth\overline{\eth})f=2\,s\,f
\end{equation}
holds. It is also important to mention that all the $\vartheta$ derivatives present in the Laplace-Beltrami operator relevant for spin $s$ fields are contained by the operator $\overline{\eth}\eth$ as the relation
\begin{equation}
\overline{\eth}\eth f=\diff_{\vartheta\vartheta} f+\cot\vartheta\,\diff_\vartheta f+\frac{1}{\sin^2\vartheta}\,\diff_{\varphi\varphi} f+2\,\mathrm{i}\,s\,\frac{\cot\vartheta}{\sin\vartheta}\,\partial_\varphi f+s\,(1-s\cot^2\vartheta)f
\end{equation}
can be seen to hold.

In addition, the spin-weighted spherical harmonics ${}_s{Y_\ell}{}^m$ are
eigenfunctions of the operators $\overline{\eth}\eth$ and $\partial_\varphi$ and their spin weight is shifted by
$\eth$ and $\overline{\eth}$,
as they satisfy
\begin{align}
\label{eq:eigb}
\eth\,{}_s{Y_\ell}^m& =\sqrt{(\ell-s)(\ell+s+1)}\cdot{}_{s+1}{Y_\ell}^m\,,\\
\overline{\eth}\,{}_s{Y_\ell}^m&=-\sqrt{(\ell+s)(\ell-s+1)}\cdot{}_{s-1}{Y_\ell}^m\,,\\
\overline{\eth}\eth\,{}_s{Y_\ell}^m&=-(\ell-s)(\ell+s+1)\cdot{}_{s}{Y_\ell}^m\,,\\
\label{eq:eige}
\partial_\varphi\,{}_s{Y_\ell}^m&=\mathrm{i} \,m\cdot{}_s{Y_\ell}^m\,.
\end{align}
Note that as $\eth\,{}_\ell{Y_\ell}^m=0$ for $s=\ell$ and  $\overline{\eth}\,{}_{-\ell}{Y_\ell}^m=0$ for $s=-\ell$,
all the spin-weighted spherical harmonics ${}_s{Y_\ell}^m$ with $\ell<|s|$ vanish.

Using these relations, it is also straightforward to check
that the spin-weighted spherical harmonics can be generated by the spin-raising $\eth$
and the spin-lowering $\overline{\eth}$ operators
starting with the conventional (zero spin-weight) spherical harmonics $Y_\ell{}^m$ as
\begin{equation}
{}_s{Y_\ell}^m=\sqrt{\frac{(\ell-s)!}{(\ell+s)!}}\cdot\eth^s{Y_\ell}^m
\end{equation}
if $0\leq s\leq \ell$ and
\begin{equation}
{}_s{Y_\ell}^m=(-1)^s\,\sqrt{\frac{(\ell+s)!}{(\ell-s)!}}\cdot\overline{\eth}^{-s}\,{Y_\ell}^m
\end{equation}
if $0>s\geq-\ell$.
Note finally that the complex conjugation acts as
\begin{equation}
\overline{{}_s{{Y}_\ell}^m}=(-1)^{s+m}{}_{-s}{Y_\ell}^{-m}\,.
\end{equation}

\medskip

In virtue of Eq. (\ref{eq:teurt}), one of the coefficients involves a multiplication by $Y_1{}^0$, whereas a division by the expression $\mathscr{A}+\mathscr{B}\cdot Y_2^0$ must also be performed.
Concerning multiplications, as it was explained in great detail in the appendixes of \cite{Csizmadia:2012kq},
once we have the expansion of scalar variables in terms of spherical harmonics,
the products of these variables can easily be evaluated by
making use of the Gaunt coefficients $G_{\ell_1}^{m_1}{}_{\ell_2}^{m_2}{}_{\ell_3}^{m_3}=\int Y_{\ell_1}^{m_1}Y_{\ell_2}^{m_2}Y_{\ell_3}^{m_3}\sin\vartheta\,\mathrm{d}\vartheta\mathrm{d}\varphi$.
Notably, completely analogous arguments apply when our spin $s$ variables are expanded
in terms of spin-weighted spherical harmonics.
The corresponding Gaunt coefficients---which now acquire three additional spin indices---can be given as
\begin{equation}
{}_{s_1s_2s_3\,}G_{\ell_1}^{\,m_1}{}_{\ell_2}^{m_2}{}_{\ell_3}^{m_3}=\int {}_{s_1}Y_{\ell_1}^{m_1}\cdot{}_{s_2}Y_{\ell_2}^{m_2}\cdot{}_{s_3}Y_{\ell_3}^{m_3}\sin\vartheta\,\mathrm{d}\vartheta\mathrm{d}\varphi\,,
\end{equation}
which can also be expressed via the Wigner-3j symbols as
\begin{multline}
{}_{s_1\,s_2\,s_3\,}G_{\ell_1}^{\,m_1}{}_{\ell_2}^{m_2}{}_{\ell_3}^{m_3}=\sqrt{\frac{(2\,\ell_1+1)(2\,\ell_2+1)(2\,\ell_3+1)}{4\pi}}\\
\times\Wj{\ell_1}{\ell_2}{\ell_3}{-s_1}{-s_2}{-s_3}\Wj{\ell_1}{\ell_2}{\ell_3}{m_1}{m_2}{m_3}\,.
\end{multline}
Then, in particular, the product $Y_{\ell_1}{}^0\cdot{}_sY_{\ell_2}{}^m$ can be evaluated as
\begin{equation}
Y_{\ell_1}{}^{0}\cdot{}_{s}Y_{\ell_2}{}^{m}=\sum_{\ell_3=|\ell_1-\ell_2|}^{\ell_1+\ell_2} (-1)^{s+m}\cdot{}_{0\,s\,-s\,}G_{\ell_1}^{\,0}{}_{\ell_2}^{m}{}_{\ell_3}^{-m}\cdot{}_{s}Y_{\ell_3}{}^{m}\,.
\end{equation}

As for the division by the term $\mathscr{A}+\mathscr{B}\cdot Y_2^0$,
note that by following the ideas introduced in \cite{Csizmadia:2012kq}
(see also \cite{Andras:2014kq, Racz:2011qu}) the division by this term can also be traced back to multiplications.
The essential observation made in \cite{Csizmadia:2012kq} was that the Neumann series expansion
\begin{equation}
A^{-1}=\sum_{k=0}^{\infty}(1-A)^k
\end{equation}
can be applied to do so. In particular, by replacing $A$ with $1+x$, where $x=\mathscr{B}/\mathscr{A}\cdot Y_2^0$,
and by choosing the value $k_{max}$ sufficiently large---as the term $\mathscr{B}/\mathscr{A}\cdot Y_2^0$
for any value of $\vartheta$ is much smaller than $1$---the approximate relation
\begin{equation}
\big[\,1+(\mathscr{B}/\mathscr{A})\cdot Y_2^0\,\big]^{-1} \approx \sum_{k=0}^{k_{max}}\ \big[\,-(\mathscr{B}/\mathscr{A})\cdot Y_2^0\,\big]^k
\end{equation}
holds. Note that in practice numerical precision with an error tolerance of $10^{-20}$
does not require the use of more than $k_{max}=12$ terms in this series.

\section{The conserved currents}
  \label{app:curr}
	\renewcommand{\theequation}{C.\arabic{equation}}
	\setcounter{equation}{0}

  In order to evaluate the energy and angular momentum balance relations we need the explicit form of $\sqrt{|h_T|}\,n^{(T)}_aE^a$, $\sqrt{|h_R|}\,n^{(R)}_aE^a$, $\sqrt{|h_T|}\,n^{(T)}_aJ^a$ and $\sqrt{|h_R|}\,n^{(R)}_aJ^a$. The following subsections list the explicit form of the implemented expressions.

  \subsection{Energy density}
  \renewcommand{\theequation}{C1.\arabic{equation}}
  \setcounter{equation}{0}

  \begin{align}
   \sqrt{|h_T|}\,n^{(T)}_aE^a=& \left(c_{TT}+c_{TTy}Y_2{}^0\right)\diff_T\Phi^{(s)}\diff_T\Phi^{(-s)}+c_{\vartheta\vartheta}\left(\Phi^{(s)}\overline{\eth}\eth\Phi^{(-s)}+\Phi^{(-s)}\overline{\eth}\eth\Phi^{(s)}\right) \nonumber \\
   &+c_{R\varphi}\left(\diff_R\Phi^{(s)}\diff_\varphi\Phi^{(-s)}+\diff_\varphi\Phi^{(s)}\diff_R\Phi^{(-s)}\right)+c_{RR}\diff_R\Phi^{(s)}\diff_R\Phi^{(-s)} \nonumber \\
   &+(c_{R}+s\, c_{Rs})\Phi^{(s)}\diff_R\Phi^{(-s)}+(c_{R}-s\, c_{Rs})\Phi^{(-s)}\diff_R\Phi^{(s)}\nonumber\\
   &+c_\varphi\left(\Phi^{(s)}\diff_\varphi\Phi^{(-s)}+\Phi^{(-s)}\diff_\varphi\Phi^{(s)}\right)+c_{0}\Phi^{(s)}\Phi^{(-s)}\,,
  \end{align}
  where
  \begin{align}
   c_{TT}=&\frac{1}{6(1 + R)R^2(1 + R^2)}\,\Big(a^2(1 + R)[-1 + 2(5 + 24M + 24M^2)R^2 \nonumber \\
   &\hskip3.0cm - (1 + 48M + 96M^2)R^4 + 48M^2R^6] + 12R[-(R(1 + R)) \nonumber \\
   & \hskip3.0cm + M(-1 - 5R - 11R^2 + R^3) + 8M^2R(-1 - 3R - R^2 + R^3) \nonumber \\
   & \hskip3.0cm + 16M^3(-1 + R)R^2(R+1)^2]\Big)\,,  \\
   c_{TTy}=&-\frac{2a^2(1 + R^2)}{3R^2}\sqrt{\frac{\pi}{5}}\,, \\
   c_{\vartheta\vartheta}=&\frac{1 + R^2}{4R^2}\,,\\
   c_{R\varphi}=&-\frac{a(R^2-1)^2}{4R^2}\,,\\
   c_{RR}=&-\frac{(R^2-1)^2(4R(R + M(R^2-1)) + a^2(R^2-1)^2)}{8R^2(1 + R^2)}\,,\\
   c_{R}=&-\frac{(R^2-1) (a^2 (R^2-1)^2 + 2 R (2 R + 2 M (R^2-1)))}{8 R^3}\,,\\
   c_{Rs}=&-\frac{(R^2-1) (4 R^2 - 2 M R  + 2 M R^3)}{8 R^3}\,,\\
   c_\varphi=&-\frac{a(-1 + R^4)}{4R^3}\,,\\
   c_{0}=&-\frac{(1 + R^2)(4R(R + M(R^2-1)) + a^2(R^2-1)^2)}{8R^4}\,.
  \end{align}

  \subsection{Energy current}
  \renewcommand{\theequation}{C2.\arabic{equation}}
  \setcounter{equation}{0}

  \begin{align}
   \sqrt{|h_R|}\,n^{(R)}_aE^a=& \ c_{TT}\,\diff_T\Phi^{(s)}\diff_T\Phi^{(-s)}+c_{TR}\left(\diff_T\Phi^{(s)}\diff_R\Phi^{(-s)}+\diff_T\Phi^{(-s)}\diff_R\Phi^{(s)}\right) \nonumber \\
   &+c_{T\varphi}\left(\diff_T\Phi^{(s)}\diff_\varphi\Phi^{(-s)}+\diff_\varphi\Phi^{(s)}\diff_T\Phi^{(-s)}\right)\nonumber\\
   &+(c_{T}+s\, c_{Ts})\Phi^{(s)}\diff_T\Phi^{(-s)}+(c_{T}-s\, c_{Ts})\Phi^{(-s)}\diff_T\Phi^{(s)}\, ,
  \end{align}
  where
  \begin{align}
   c_{TT}=&\frac{1}{R+R^3}(a^2(-1 + 2M(R^2-1))(R^2-1)^2 \nonumber \\ &+ 2(-2R^2 + 4M^2R(R^2-1)^2 + M(1 + 4R + 3R^2)(R-1)^2))\,,\\
   c_{TR}=&\frac{(R^2-1)^2(4R(R + M(R^2-1)) + a^2(R^2-1)^2))}{8R^2(1 + R^2)}\,,\\
   c_{T\varphi}=&\frac{a(R^2-1)^2}{4R^2}\,,\\
   c_{T}=&\frac{(R^2-1) (a^2 (R^2-1)^2 + 2 R (2 R + 2 M (R^2-1)))}{8 R^3}\,,\\
   c_{Ts}=&\frac{(R^2-1) (4 R^2 -2 M R + 2 M R^3)}{8 R^3}\,.
  \end{align}

  \subsection{Angular momentum density}
  \renewcommand{\theequation}{C3.\arabic{equation}}
  \setcounter{equation}{0}

  \begin{align}
   \sqrt{|h_T|}\,n^{(T)}_aJ^a=& \
   \left(c_{T\varphi}+c_{T\varphi y}Y_2{}^0\right)\left(\diff_T\Phi^{(s)}\diff_\varphi\Phi^{(-s)}+\diff_\varphi\Phi^{(s)}\diff_T\Phi^{(-s)}\right) \nonumber \\
   &+c_{R\varphi}\left(\diff_R\Phi^{(s)}\diff_\varphi\Phi^{(-s)}+\diff_\varphi\Phi^{(-s)}\diff_R\Phi^{(s)}\right)+c_{\varphi\varphi}\diff_\varphi\Phi^{(s)}\diff_\varphi\Phi^{(-s)} \nonumber  \\
   &+\left(c_{\varphi}+s\,(c_{\varphi s}+c_{\varphi y}\mathrm{i} Y_1{}^0)\right)\Phi^{(s)}\diff_\varphi\Phi^{(-s)}\nonumber\\
   &+\left(c_{\varphi}-s\,(c_{\varphi s}+c_{\varphi y}\mathrm{i} Y_1{}^0)\right)\Phi^{(-s)}\diff_\varphi\Phi^{(s)}\,,
  \end{align}
  where
  \begin{align}
   c_{R\varphi}=&\frac{1}{2(R+R^3)}\,\Big(a^2(-1 + 2M(R^2-1))(R^2-1)^2 \nonumber \\& \hskip2.8cm + 2(-2R^2 + 4M^2R(R^2-1)^2 + M(1 + 4R + 3R^2)(R-1)^2)\Big)\,,\\
   c_{T\varphi}=&\frac{1}{6(1 + R)R^2(1 + R^2)}\Big(a^2(1 + R)(-1 + 2(5 + 24M + 24M^2)R^2  \nonumber \\
    &\hskip4.0cm - (1 + 48M + 96M^2)R^4 + 48M^2R^6) + 12R[-R(1 + R)  \nonumber \\
    & \hskip3.2cm + M(-1 - 5R - 11R^2 + R^3) + 8M^2R(-1 - 3R - R^2 + R^3)  \nonumber \\& \hskip4.2cm + 16M^3(-1 + R)R^2(R+1)^2]\Big)\,,\\
   c_{T\varphi y}=&-\frac{2a^2(1 + R^2)}{3R^2}\sqrt{\frac{\pi}{5}}\,,\\
   c_{\varphi\varphi}=&\frac{2a(-1 + 2M(R^2-1))}{R} 
   \end{align}
   \begin{align}
   c_{\varphi}=&\frac{1}{2 R^2 (R^2-1)}\left(a^2 (R^2-1)^2 (2 M (R^2-1)-1)\right.\nonumber\\
   &\left.+ 2 (4 M^2 R (R^2-1)^2 -2 R^2 + M (R-1)^2 (1 + 4 R + 3 R^2))\right)\,,\\
   c_{\varphi s}=&\frac{\left(4 M^2 (R-1) R (R+1)^2+M \left(7 R^3+5 R^2-3 R-1\right)+2 (R-1) R\right)}{2 R^2 (R+1)}\,,\\
   c_{\varphi y}=&-\frac{ a \left(R^2+1\right)}{R^2}\sqrt{\frac{\pi }{3}}\,.
  \end{align}

  \subsection{Angular momentum current}
  \renewcommand{\theequation}{C4.\arabic{equation}}
  \setcounter{equation}{0}

  \begin{align}
   \sqrt{|h_R|}\,n^{(R)}_aJ^a=& \
   c_{R\varphi}\left(\diff_R\Phi^{(s)}\diff_\varphi\Phi^{(-s)}+\diff_R\Phi^{(-s)}\diff_\varphi\Phi^{(s)}\right)+c_{\varphi\varphi}\diff_\varphi\Phi^{(s)}\diff_\varphi\Phi^{(-s)} \nonumber \\
   &+c_{T\varphi}\left(\diff_T\Phi^{(s)}\diff_\varphi\Phi^{(-s)}+\diff_\varphi\Phi^{(s)}\diff_T\Phi^{(-s)}\right) \nonumber \\
   &+(c_{\varphi}+s\, c_{\varphi s})\Phi^{(s)}\diff_\varphi\Phi^{(-s)}+(c_{\varphi}-s\, c_{\varphi s})\Phi^{(-s)}\diff_\varphi\Phi^{(s)}\,,
  \end{align}
  where
  \begin{align}
   c_{R\varphi}=&\frac{(R^2-1)^2(4R(R + M(R^2-1)) + a^2(R^2-1)^2)}{8R^2(1 + R^2)}\,,\\
   c_{T\varphi}=&\frac{1}{2(R+R^3)}\Big(a^2(-1 + 2M(R^2-1))(R^2-1)^2 + 2(-2R^2 + 4M^2R(R^2-1)^2 \nonumber \\&+ M(1 + 4R + 3R^2)(R-1)^2)\Big)\,, \\
   c_{\varphi\varphi}=&\frac{a(R^2-1)^2}{2R^2}\,,\\
   c_{\varphi}=&\frac{\left(R^2-1\right) \left(a^2 \left(R^2-1\right)^2+2 R \left(2 M \left(R^2-1\right)+2 R\right)\right)}{8 R^3}\,,\\
   c_{\varphi s}=&\frac{\left(R^2-1\right) \left(2 M R^3-2 M R+4 R^2\right)}{8 R^3}\,.
  \end{align}





\end{document}